\documentclass[preprint,11pt,3p]{elsarticle}

\usepackage{graphicx}
\usepackage[dvipsnames]{xcolor}
\usepackage{epstopdf, epsfig}
\usepackage{color,soul}
\usepackage{physics}
\usepackage{multirow}
\usepackage[version=4]{mhchem}
\usepackage{caption}
\usepackage{mwe}
\usepackage{longtable,tabularx}
\usepackage{mathtools}
\usepackage{amsmath,mismath}
\usepackage{amssymb}

\usepackage{hyperref}
\hypersetup{
    colorlinks=true,
    linkcolor=blue,
    filecolor=magenta,      
    urlcolor=cyan,
}

\newcommand{\Rey}{\mathit{Re}}

 \usepackage{lineno,hyperref}

\biboptions{authoryear}

\makeatletter
\def\ps@pprintTitle{%
 \let\@oddhead\@empty
 \let\@evenhead\@empty
 \def\@oddfoot{}%
 \let\@evenfoot\@oddfoot}
\makeatother

\begin{document}

\begin{frontmatter}

\title{Gappy spectral proper orthogonal decomposition}

\author{Akhil Nekkanti}
\ead{aknekkan@eng,ucsd.edu}
\author{Oliver T. Schmidt\corref{cor1}}
\ead{oschmidt@ucsd.edu}
\address{Department of Mechanical and Aerospace
  Engineering, University of California San Diego, La Jolla, CA, USA}
\cortext[cor1]{Corresponding author}



\begin{abstract}
Experimental spatio-temporal flow data often contain gaps or other types of undesired artifacts. To reconstruct flow data in the compromised or missing regions, a data completion method based on spectral proper orthogonal decomposition (SPOD) is developed. The algorithm leverages the temporal correlation of the SPOD modes with preceding and succeeding snapshots, and their spatial correlation with the surrounding data at the same time instant. For each gap, the algorithm first computes the SPOD of the remaining, unaffected data. In the next step, the compromised data are projected onto the basis of the SPOD modes. This corresponds to a local inversion of the SPOD problem and yields expansion coefficients that permit the reconstruction in the affected regions. This local reconstruction is successively applied to each gap. After all gaps are filled in, the procedure is repeated in an iterative manner until convergence. This method is demonstrated on two examples: direct numerical simulation of laminar flow around a cylinder, and time-resolved PIV data of turbulent cavity flow obtained by \citet{zhang2019spectral}. Randomly added gaps correspond to 1\%, 5\%, and 20\% of data loss. Even for 20\% data corruption, and in the presence of measurement noise in the experimental data, the algorithm recovers 97\% and 80\% of the original data in the corrupted regions of the simulation and PIV data, respectively. These values are higher than those achieved by established methods like gappy POD and Kriging.
\end{abstract}


 \end{frontmatter}


\section{Introduction}

Gappy data reconstruction techniques find wide use in the completion of partially missing or otherwise compromised experimental data. The focus of this work is the estimation of missing regions in long time series of spatially resolved, generally turbulent, statistically stationary flow data. One of the most widely used experimental technique to acquire such data for turbulence research \citep{westerweel2013particle} and geophysical flows \citep{doron2001turbulence}, but also in the automotive \citep{beaudoin2008drag,conan2011experimental} and aerospace industries \citep{willert1997stereoscopic}, is particle image velocimetry (PIV, see e.g. \cite{raffel1998particle,adrian2011particle}). Missing or corrupted regions in PIV measurements have many sources. Among them are shadowing, that is, the partial obstruction of the laser sheet, reflections from objects, and the inaccessibility of certain regions for the imaging system \citep{westerweel1994efficient,huang1997errors,hart2000piv,sciacchitano2012navier}. Corrupted regions also arise from irregular seeding or the absence of a sufficient number of tracer particles, for example, in recirculation zones. Whereas the latter sources of errors are specific to PIV, other measurement techniques suffer from similar problems. Atmospheric data obtained via satellite imagery, for example, often suffer from partial obstruction due to cloud coverage \citep{alvera2005reconstruction}.

Standard mathematical techniques for approximating missing data include basic interpolation and least-square estimation. Techniques devised specifically for the task of gappy data reconstruction include optimal interpolation \citep{reynolds1994improved,smith1996reconstruction,kaplan1998analyses} and Kriging \citep{oliver1990kriging,stein1999interpolation}. Kriging, the more popular method, uses a local or global interpolant that is evaluated at the missing points based on local weighted averaging. An advantage of Kriging is that it is inherently capable of extrapolation, i.e, it can reconstruct spatial gaps that are persistent in time. The use of proper orthogonal decomposition (POD) in conjunction with least-squares estimation for data reconstruction was proposed by \citet{everson1995karhunen}. This original `gappy POD' algorithm was later extended by \citet{venturi2004gappy} and shown to outperform Kriging for the reconstruction of flow data with up to $50\%$ of gappyness.  A computational efficient algorithm that does not require the repeated solutions of least-squares problems was introduced by \citet{gunes2006gappy}. Since its introduction, gappy POD has been applied to various flows such as the flow past an airfoil \citep{bui2004aerodynamic,willcox2006unsteady},  cavity flow \citep{murray2007application}, boundary layers \citep{gunes2007spatial,raben2012adaptive}, gas turbine combustors \citep{saini2016development}, and arterial blood flow \cite{yakhot2007reconstruction}. It has also become an essential component of several model reduction strategies that use POD modes as their basis \citep{chaturantabut2010nonlinear,benner2015survey}. Another application is efficient sensor placement \citep{willcox2006unsteady,yildirim2009efficient}. In the ocean sciences, gappy POD was independently developed under the name of data interpolating empirical orthogonal functions (DINEOF) by \citet{beckers2003eof} and was used to, for example,  fill in observational sea surface temperature \citep{alvera2005reconstruction,beckers2006dineof}, Southern Oscillation Index \citep{kondrashov2006spatio}, and surface chlorophyll concentration data \citep{alvera2007multivariate,taylor2013sensitivity}.

In this work, we propose an algorithm based on frequency-domain, or spectral proper orthogonal decomposition (SPOD), for gappy data reconstruction. Even though the mathematical foundations of SPOD were first laid out by \citet{lumley1970stochastic}, the method was rarely used, with a few notable exceptions like the work of \citet{glauser1987coherent}, before its first application to very large numerical data by \citet{schmidt2018spectral}. In the same year, \citet{towne2018spectral} established relationships to hydrodynamic stability theory and other modal decompositions. The proposed `gappy SPOD' algorithm is fundamentally different from those of \citet{everson1995karhunen}  and \citet{venturi2004gappy}, both algorithmically and in that it leverages both spatial and temporal coherence. Note that space-only POD modes are only correlated at zero time lag, and hence do not possess the latter property \citep{towne2018spectral}.


The paper is structured as follows. A brief overview of SPOD and SPOD-based reconstruction is given in \S \ref{methodology}. \S \ref{algorithm} describes the gappy SPOD algorithm. In \S\S \ref{example 1} and \ref{example 2}, we demonstrate the algorithm on numerical simulation data of laminar cylinder flow and PIV data of turbulent cavity flow, respectively. \S \ref{Comparison} compares the performance of the proposed gappy SPOD algorithm to established gappy POD and Kriging methods. Finally, \S \ref{conclusion} summarizes this work. The performance for a case involving sequences of missing snapshots is analysed in \ref{Appendix: miss snapshots}. Parameter studies on the effect of the spectral estimation parameters, as well as data length and SPOD convergence are presented in \ref{Appendix: nfft}, and \ref{Appendix: convergence}, respectively.




\section{Methodology} \label{methodology}
\subsection{Spectral Proper Orthogonal Decomposition}
SPOD extracts monochromatic modes that optimally capture, depending on the choice of norm, the flow's energy. These modes are computed as the eigenvectors of the cross-spectral density matrix. We use a specific SPOD algorithm that estimates the cross-spectral density matrix based on Welch’s approach \citep{welch1967use}. The implementation of this approach entails partitioning the entire time series into smaller segments that are interpreted as independent realizations of the flow. Next, we briefly introduce the notation used in this work. For a detailed mathematical derivation and the algorithmic implementation, the reader may refer to, for example, \citet{towne2018spectral} or \citet{schmidt2020guide}.  


We denote by 
\begin{equation}
\vb{q}_i = \vb{q}(t_i), \qquad i=1,2,\cdots, n_t,
\end{equation}
the ensemble of $n_t$ snapshots of the statistically stationary flow field with its mean removed. The first step of the standard Welch approach is to segment the data into $n_{\rm blk}$ overlapping blocks, each containing $n_{\rm fft}$ snapshots. If the blocks overlap by $n_{\rm ovlp}$ snapshots, then the $j$-th column in the $k$-th block is given by $\vb{q}_{j}^{(k)}=\vb{q}_{j+(k-1)(n_{\rm fft}-n_{\rm ovlp})+1}$, where $k = 1,2,\cdots, n_\textrm{blk}$, and $j = 1,2,\cdots, n_\textrm{fft}$. Next, we compute a windowed temporal discrete Fourier transform and arrange all the Fourier realizations of the $l$-th frequency, $\Hat{\vb{q}}_{l}^{(j)}$, into a matrix, 
\begin{equation}
\Hat{\vb{Q}}_{l}=\bqty{\Hat{\vb{q}}_{l}^{(1)}, \Hat{\vb{q}}_{l}^{(2)}, \cdots, \Hat{\vb{q}}_{l}^{(n_{\rm blk})} }.
\label{eq 8}
\end{equation}
The SPOD modes, $\vb*{\Phi}_l$, and eigenvalues, $\vb*{\Lambda}_l$, are finally obtained as the eigenpairs of the weighted CSD matrix $\vb{S}_l= \frac{1}{n_\textrm{blk}}\Hat{\vb{Q}}_{l} \Hat{\vb{Q}}_{l}^{*}\vb{W}$, where $\vb{W}$ is a positive-definite Hermitian matrix that accounts for the component-wise and numerical quadrature weights. For data with more spatial degrees of freedom than the number of snapshots, we solve the smaller eigenvalue problem, 
\begin{equation}
\frac{1}{n_{\rm blk}}\vb{\Hat{Q}}_{l}^{*}\vb{W} \vb{\Hat{Q}}_{l} \vb*{\Psi}_{l}=\vb*{\Psi}_{l} \vb*{\Lambda}_{l},
\label{eq 11}
\end{equation}
for the (unscaled) expansion coefficients, $\vb*{\Psi}_{l}$. The SPOD modes at the $l$-th frequency are recovered as 
\begin{equation}
\vb*{\Phi}_{l} = \frac{1}{\sqrt{n_\textrm{blk}}}\vb{\Hat{Q}}_{l} \vb*{\Psi}_{l} \vb*{\Lambda}_{l}^{-1/2}.
\label{eq 12}
\end{equation}
 The column vectors of $\vb*{\Phi}_{l}=[ \vb*{\phi}_{l}^{(1)}, \vb*{\phi}_{l}^{(2)}, \cdots , \vb*{\phi}_{l}^{(n_{\rm blk})} ] $ are the SPOD modes and the diagonal entries of $\vb*{\Lambda}_{l}=\text{diag} ( \lambda_{l}^{(1)}, \lambda_{l}^{(2)}, \cdots , \lambda_{l}^{(n_{\rm blk})} ) $ contain the mode energies.  An important property of the SPOD modes is their orthogonality in their weighted inner product, $\big<\vb*{\phi}_{l}^{(i)},\vb*{\phi}_{l}^{(j)}\big>=\vb*{\phi}_{l}^{(i)}\vb{W} \vb*{\phi}_{l}^{(j)}= \delta_{ij}$, at each frequency. The associated norm is denoted by $\|\cdot\|_2$. 

\subsection{Data reconstruction}
The reconstruction of the original data is based on the inversion of the SPOD problem, comprehensively discussed in \citet{nekkanti2021frequency}.  The Fourier realizations at each frequency are reconstructed from the SPOD modes as  $\vb{\Hat{Q}}_l = \vb*{\Phi}_l\vb{A}_l$. Here, $\vb{A}_l$ is the matrix of the (scaled) expansion coefficients computed as 
\begin{align}
        \vb{A}_l &= \sqrt{n_{\rm blk}} \vb*{\Lambda}^{1/2}_{l} \vb*{\Psi}^*_{l},  \quad \text{or} \label{expansion coefficients 1} \\   
        \vb{A}_l &= \vb{\boldsymbol{\Phi}}_l^{*} \vb{W} \Hat{\vb{Q}}_{l}. \label{expansion coefficients 2}
\end{align}
The expansion coefficients can be saved during the computation of SPOD using equation (\ref{expansion coefficients 1}) or recovered later by projecting the Fourier realizations onto the modes using equation (\ref{expansion coefficients 2}). Using the expansion coefficients $a_{ik}$ contained in $\vb{A}$ at any given frequency, the $k$-th block can be reconstructed as 
\begin{equation}
\vb{Q}^{(k)}=\mathcal{F}^{-1} \bqty{ \pqty{\sum\limits_{i} a_{ik} \vb*{\phi}^{(i)}}_{\!\!\!l=1} , \pqty{\sum\limits_{i} a_{ik} \vb*{\phi}^{(i)}}_{\!\!\!l=2} , \cdots, \pqty{\sum\limits_{i} a_{i k} \vb*{\phi}^{(i)}}_{\!\!\!l=n_{\rm fft}}}, 
\label{reconstruction blocks}
\end{equation}
where $\mathcal{F}^{-1} $ is the inverse windowed Fourier transform. Finally, the time series is reconstructed from the data segments by computing the average of the reconstructions from overlapping blocks, weighted by the relative value of their windowing function \citep{nekkanti2021frequency}.

\section{Algorithm} \label{algorithm}

The gappy SPOD algorithm consists of three loops: in the local loop, a single gap is converged by repeated reconstruction with continuously updated local expansion coefficients; in the inner loop, this process is repeated for all gaps; in the outer loop, the global convergence of the inner loops is assessed.\newline

\noindent{\bf Algorithm:} Gappy SPOD
\begin{enumerate}[(i)]
    \item  Segment the time series into overlapping blocks and compute the temporal Fourier transform of each block (if not computed in the previous iteration).
    \item Proceed to the $n$-th gap and choose all the realizations of the Fourier transform that are \emph{not} affected by this gap.
    \item Compute the SPOD from all the Fourier realizations that are not affected by this gap (equations (\ref{eq 11}) and (\ref{eq 12})) and store the corresponding expansion coefficients (equation (\ref{expansion coefficients 1})).  (For denoising, apply mode truncation; see \ref{Appendix: noise_added}.)
    \item Compute the SPOD expansion coefficients for blocks \emph{affected} by the $n$-th gap by projecting their Fourier transforms onto the SPOD basis (equation (\ref{expansion coefficients 2})). 
    \item Reconstruct the affected blocks by inverting the SPOD (equation (\ref{reconstruction blocks})) from the expansion coefficients computed in (iii) and (iv); replace the regions affected by the $n$-th gap.
    \item (local loop) Go to (iv) to update the expansion coefficients now that the data is reconstructed in the affected regions until the convergence criterion based on the change of reconstruction of the $n$-th gap is met. 
    \item (inner loop) Set $n\leftarrow n+1$ and go to (i) until all gaps are reconstructed.
    \item (outer loop) Set $n = 1$ and go to (i); repeat until convergence criterion based on change of reconstruction between outer loop iterations is met.
\end{enumerate}

The metrics used to gauge the error, as compared to the true solution, and the convergence for the local and outer loops, as compared to the previous iteration within the respective loop, are defined next. 

\subsection{Error and Convergence metrics}
Define $G$ as the index set corresponding to all gappy snapshots and $G_n \subset G$ as the subset of indices corresponding to the $n$-th gap. The following error and convergence metrics are used to evaluate the efficacy of our method:
    \begin{align}
        e_i&= \frac{\Big\|\vb{q}^{g}_{i}-\vb{\tilde{q}}^{g}_{i}\Big\|_2^2}{\Big\|\vb{q}^{g}_{i}\Big\|_2^2} \quad \text{(relative error of $i$-th snapshot),}  \label{error}\\
        e_n&= \sum\limits_{i \in G_n } e_i \quad \text{(relative error of $n$-th gap),}  \label{error}\\
        E &= \sum\limits_{i \in G } e_i \quad \text{(global relative error),} \label{global relative error}\\
        c_n &=\sum\limits_{i \in G_n} \frac{\Big\|\vb{\tilde{q}}^{{g}^{[j-1]}}_{i}-\vb{\tilde{q}}^{g^{[j]}}_{i}\Big\|_2^2}{\Big\|\vb{\tilde{q}}^{g^{[j-1]}}_{i}\Big\|_2^2} \quad \text{(convergence of $n$-th gap).} \label{convergence}
    \end{align}
    Here, $\vb{q}^{g}$ and $\vb{\tilde{q}}^{g}$ are, respectively, the original and reconstructed data in the gappy regions only, and superscript $[j]$ the iteration index. The calculation of the relative errors, equations (\ref{error}) and (\ref{global relative error}), requires knowledge of the original data, $\vb{q}$.  For demonstration purposes only, artificial gaps were introduced in this study, and hence the relative errors can be computed. The convergence metric defined in equation (\ref{convergence}) does not require the true data and can be evaluated even for gappy datasets. Throughout this paper, the thresholds used for convergence criteria are $\mathit{tol} = 10^{-8}$. We will demonstrate later that this tolerance is very conservative in that the error is long converged before the criterion is met for our examples.  
     
    \subsection{Computational complexity}        
 The computational cost of gappy SPOD is dominated by the matrix multiplications in the local loops, i.e, step (vi) of the algorithm in \S \ref{algorithm}. Welch's algorithm outlined in \S 2 segments the dataset of size into $n_{\rm{blk}}$ blocks. Here $m$ is the product of the spatial degrees of freedom and the number of variables. For flow data with a high spatial resolution, as the two cases considered here, we can typically assume that $m \gg n_{\rm{fft}},n_{\rm{blk}}$. The SPOD in step (iii) requires the formation of the cross-spectral density matrix, its eigenvalue decomposition, and the computation of the SPOD modes at each frequency. The time complexities of these operations are $\bigo(m n_{\rm{blk}}^2)$, $\bigo(n_{\rm{blk}}^3)$, and $\bigo(m n_{\rm{blk}}^2)$, respectively. The total time complexity of step (iii) is hence $\bigo( m n_{\rm{fft}} n_{\rm{blk}}^2)$. The time complexities of steps (iv) and (v) are $\bigo(m n_{\rm{fft}} n_{\rm{blk}} n_{\rm{blk,gaps}})$,  and $\bigo(m n_{\rm{fft}}  n_{\rm{blk,gaps}} \log(n_{\rm{blk,gaps}}))$, respectively, where $n_{\rm{blk,gaps}}$ is the number of blocks that contain gaps. The number of local loops required for the convergence of each gap is an  \textit{a priori} unknown number $n_{\rm{local}}$. As each inner loop converges all gaps, which amounts to a time complexity of  $\bigo(m n_{\rm{gaps}} n_{\rm{fft}} n_{\rm{blk}} (n_{\rm{blk}}  + n_{\rm{local}} n_{\rm{blk,gaps}} )) $. In practice, we find that $n_{\rm{local}} n_{\rm{blk,gaps}} \gg  n_{\rm{blk}}$ and the estimate simplifies to $\bigo(m n_{\rm{gaps}} n_{\rm{fft}} n_{\rm{blk}} n_{\rm{local}} n_{\rm{blk,gaps}}) $. Finally, if the algorithm requires $n_{\rm{outer}}$ outer loops for convergence, then its total time complexity is $\bigo(m n_{\rm{out}} n_{\rm{local}} n_{\rm{gaps}} n_{\rm{fft}} n_{\rm{blk}} n_{\rm{blk,gaps}} )$. 

\section{Results}

 \begin{table*}
  \centering
    \begin{tabular}{ccccccccccc}
    \hline
{Case} & {Variables} &{$\Rey$} & {$n_x$} & {$n_y$} & {$n_z$} & {$n_t$} & {$ \Delta t$} &{$n_{\rm{fft}}$} & {$n_{\rm{ovlp}}$} & {$n_{\rm{blk}}$} \vspace{0.1em}  \\ \cline{1-11} \vspace{-0.5em}\\
 Cylinder DNS                      & $u$,$v$    & 500 & 250 & 125 & 1 & 4096 & 0.06 &256 & 128 & 31 \vspace{0.2em} \\ 
Cavity PIV                        & $u$,$v$     & 3.3 $\times 10^{5}$ & 150 & 55 & 1 & 16000 & $6.25\times 10^{-5}$s &256 & 128 & 124 \vspace{0.2em} \\\cline{1-11}
\end{tabular}
\caption{Parameters example databases and spectral estimation parameters.}
\end{table*}

We demonstrate the reconstruction of missing data using the gappy SPOD reconstruction method on two examples. The parameters of these example datasets are given in table 1. The first example is direct numerical simulation (DNS) data of the canonical laminar flow past a cylinder. This flow is a popular benchmark case that we will use to compare with established methods such as gappy POD \citep{everson1995karhunen,venturi2004gappy} and Kriging \citep{oliver1990kriging}. The second example is the experimental data of a high Reynolds number flow over an open cavity obtained from time-resolved particle image velocimetry (PIV) measurements \citep{zhang2017identification,zhang2020spectral}. This data exemplifies realistic turbulent flow data and is subject to measurement noise.

\subsection{Example 1: Cylinder flow at $\Rey$=500} \label{example 1}

\begin{figure}[!h]
\centering
{\includegraphics[trim={0cm 4.4cm 0cm 0.9cm },clip,width=0.9\textwidth]{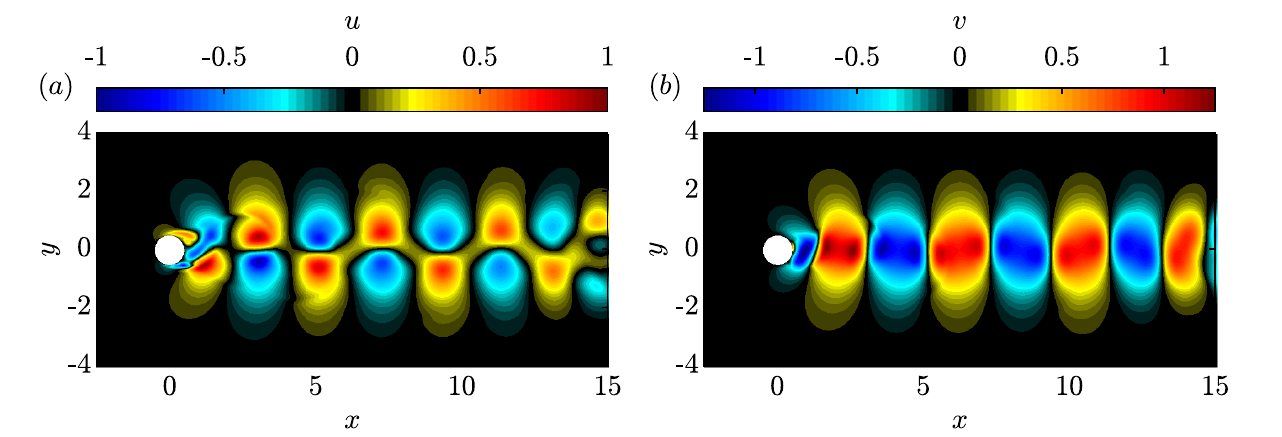}}
\caption{Instantaneous fluctuating flow field for flow around a cylinder at $\Rey = 500$: ($a$) streamwise
velocity, $u$; ($b$) transverse velocity, $v$.}
\label{fig 3.0}
\end{figure}

As a first example, we consider the flow around a cylinder at a Reynolds number $\Rey=\rho U_{\infty} D/\mu = 500$, where $\rho$ is the density, $U_{\infty}$ the incoming flow velocity, $\mu$ the dynamic viscosity, and $D$ the diameter of the cylinder. The coordinates are non-dimensionalized by $D$ and velocities by $U_\infty$, respectively. This data was obtained by solving the two-dimensional incompressible Navier-Stokes equations for the velocity field $\vb{q}= \bqty{u, v}^{T}$ using the immersed-boundary solver by \citet{goza2017strongly}. The data was acquired after all transients have subsided. Instantaneous flow fields of the streamwise velocity, $u$, and transverse velocity, $v$, are shown in figure \ref{fig 3.0}.

\begin{figure}[!tp]
\centering
{\includegraphics[trim={0cm 1.75cm 2.25cm 0.25cm },clip,width=0.9\textwidth]{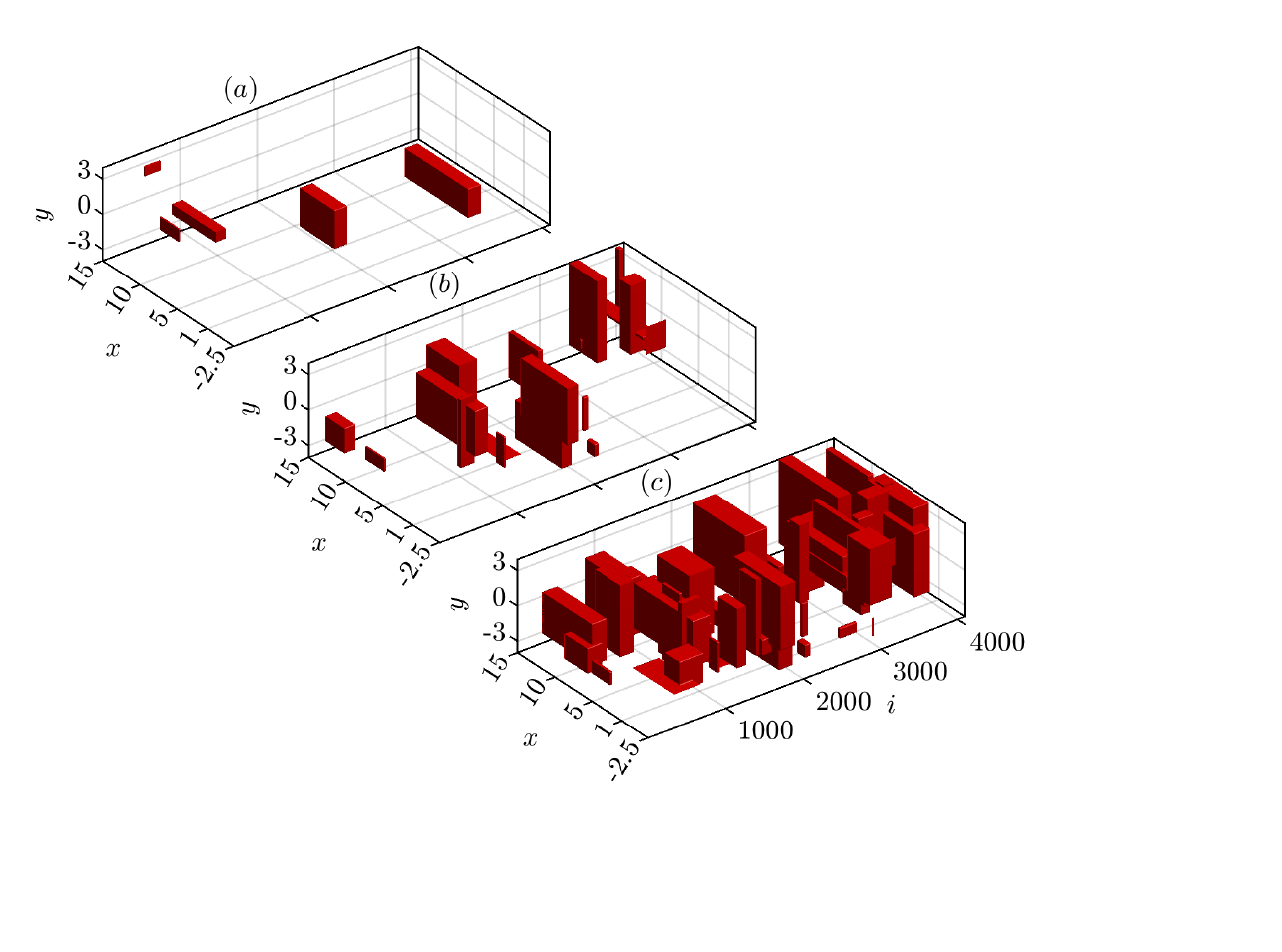}}
\caption{Randomly generated gaps in the flow past a cylinder: (a) 1\% gappyness; (b) 5\% gappyness; (c) 20\% gappyness.  Red blocks indicate gaps. The streamwise ($x$-$y$) plane is plotted over the snapshot index, $i$.}
\label{fig 3.1}
\end{figure}

The gappy SPOD reconstruction is demonstrated in three scenarios with 1\%, 5\%, and 20\% of missing data. Missing regions, or gaps, are artificially introduced to the data after the mean is removed. The gaps are randomly seeded in space and time. Similarly, the spatial and temporal extent of the gaps is randomly sampled. Figure \ref{fig 3.1} shows the spatio-temporal distribution of the gaps for all three levels of gappyness. The gaps are allowed to extend over the entire field-of-view and up to 300 snapshots. The gappy-SPOD algorithm reconstructs the data by filling gaps in sequential order. Here, we number the gaps according to their first occurrence in time and have verified that the final error upon convergence is insensitive to the order by which the algorithm handles the gaps. Figure \ref{fig 3.1}($c$) shows the extreme example of 20\% gappyness in which every block contains missing data. Following best practices for SPOD \citep{schmidt2020guide}, the data is segmented into 31 blocks with 256 snapshots and 50\% overlap. The effect of the parameter $n_\textrm{fft}$ is investigated in \ref{Appendix: nfft}. Additionally, in \ref{Appendix: miss snapshots}, we explore a case with 5\% missing data and entire snapshots, or short sequences of snapshots, missing from the data.

\begin{figure}[!tp]
\centering
{\includegraphics[trim={0cm 0.88cm 0cm 0.15cm },clip,width=1.0\textwidth]{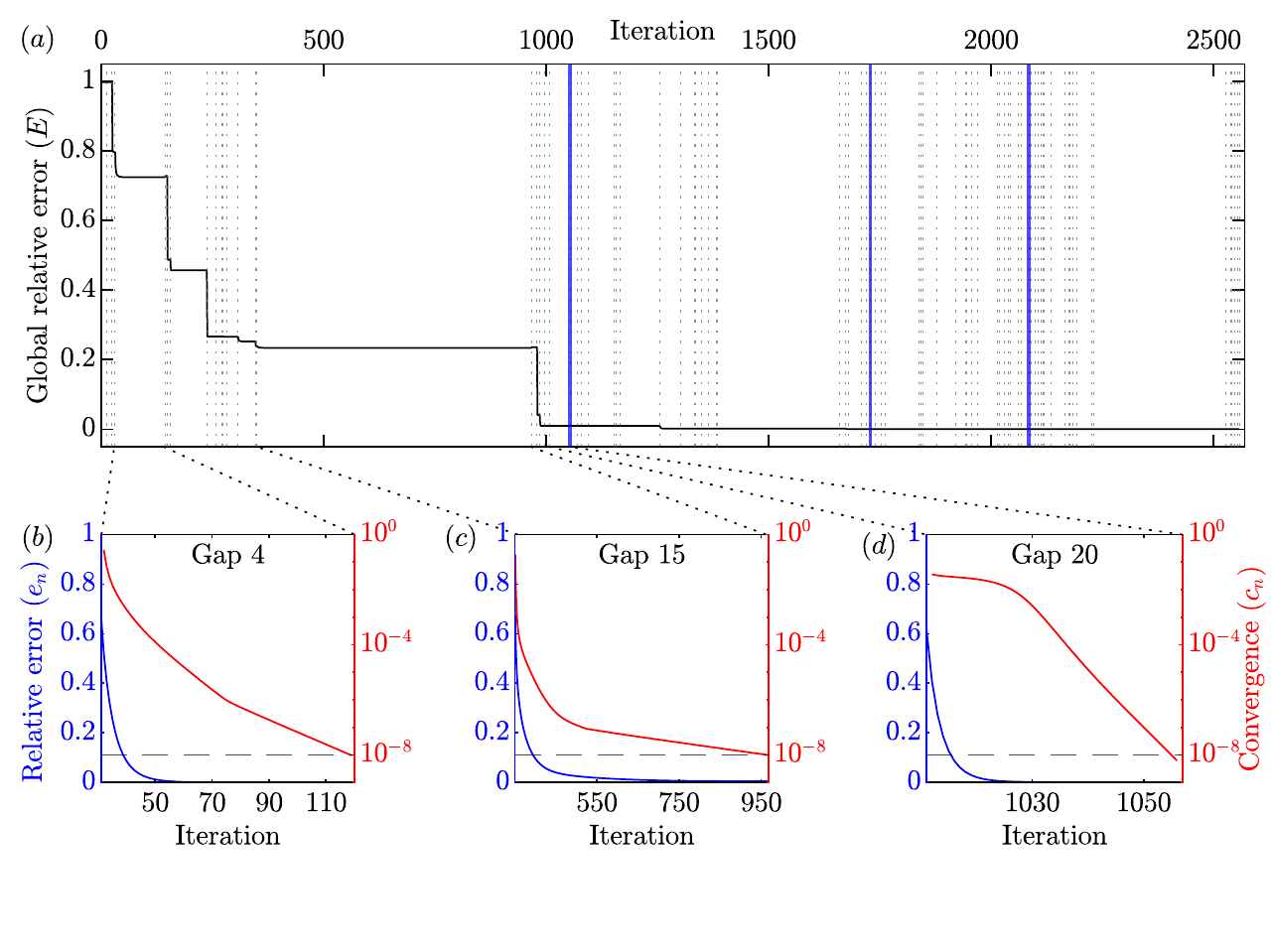}}
\caption{Errors and convergence for 5\% gappyness, see figure \ref{fig 3.1}($b$). The global relative error is shown in ($a$). Inner (gap-wise) iterations are denoted by grey dotted lines and outer iterations (sweeps over all gaps) as blue solid lines. Panels ($b$-$d$) show the gap-wise error and convergence for three randomly selected gaps within the first outer loop.}
\label{fig 3.2}
\end{figure}

We start by exploring the 5\% gappyness case, previously shown in figure 2($b$). This case consists of $n_\mathrm{gaps}=20$ randomly seeded and sized gaps. Figure \ref{fig 3.2} illustrates the local and global error s and the convergence of the algorithm. By `local', we refer to the gap-wise iteration, that is, steps (iv)-(vi) of the algorithm in \S \ref{algorithm}, and by `global' to the outer iteration loop over steps (i)-(viii). Figure \ref{fig 3.2}($a$) shows the evolution of the global relative error as defined in equation (\ref{global relative error}) as all gaps are converged to $c_n \leq \mathit{tol}=1\times10^{-8}$ within each of the four outer loops. The gap-wise convergence, $c_n$, is defined by equation (\ref{convergence}). The four outer loops required to achieve global convergence are marked by the thick blue lines. Each of these outer iterations comprises $n_\mathrm{gaps}$ inner loops, indicated by the grey dotted lines. The global relative error is normalized according to equation (\ref{global relative error}), and hence starts with one. The error non-strictly monotonically decreases as the gaps are sequentially filled in. This is a desirable property, but we note that the algorithm does not guarantee it. The amount by which each inner loop can reduce the global relative error is dependent on the spatial location and temporal extent of the corresponding gap. This explains the sharp drops at the beginning of some local iteration loops.  At the end of the first outer loop, the global relative error is already reduced to 1.3\%, and its final value after four outer loops is $7.7 \times 10^{-5}$. Notably, a large fraction of the reconstruction error reduction is accomplished by the first outer loop. A quantitative comparison with other methods such as the gappy-POD method by \citet{venturi2004gappy} is provided in section \S \ref{Comparison}.

Panels \ref{fig 3.2}($b$-$d$) show the local relative errors and convergence for three representative gaps during the first outer loop. The local relative error for all gaps decreases monotonically from one to very small values of order $10^{-3}$. After the final outer iteration loop, the local relative errors are of order $10^{-5}$. The tolerance of $\mathit{tol} =1\times10^{-8}$, as indicated by blacked dash lines, guarantees that the relative errors are all well converged.

\begin{figure}[!tp]
\centering
{\includegraphics[trim={0cm 3.45cm 0cm 0.76cm },clip,width=1.0\textwidth]{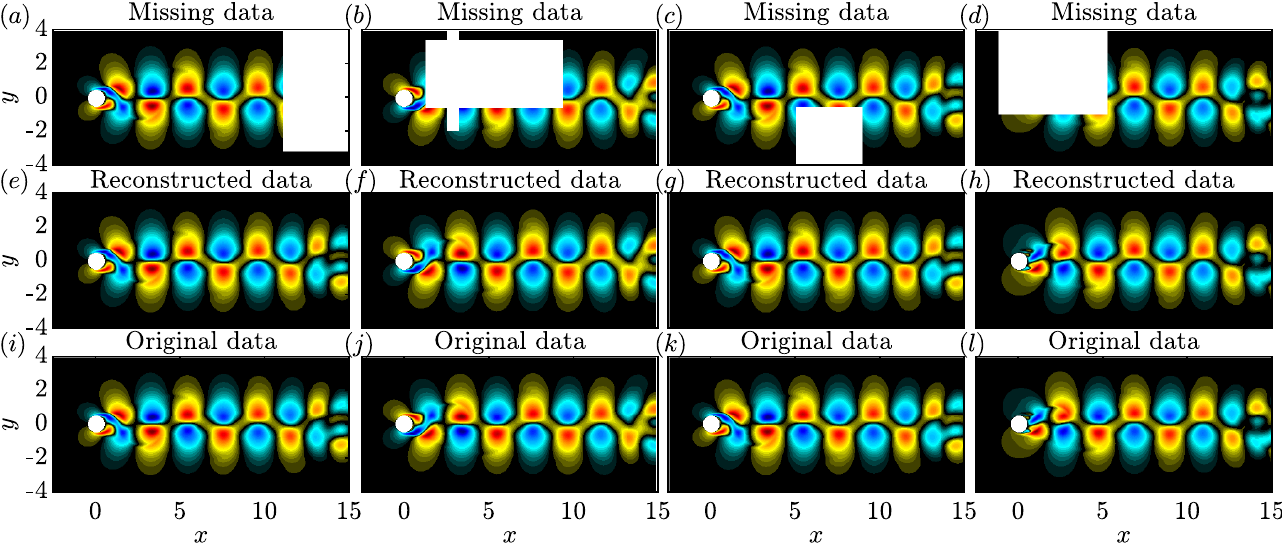}}
\caption{Reconstruction for 5\% gappyness at four time instances:  ($a$-$d$) gappy data; ($e$-$h$) reconstructed data; ($i$-$l$) original data. False colors of the streamwise velocity fluctuations, $u$, are on the same scale in all plots. }
\label{fig_5_per_u_vis}
\end{figure}

\begin{figure}[!tp]
\centering
{\includegraphics[trim={0cm 3.45cm 0cm 0.76cm },clip,width=1.0\textwidth]{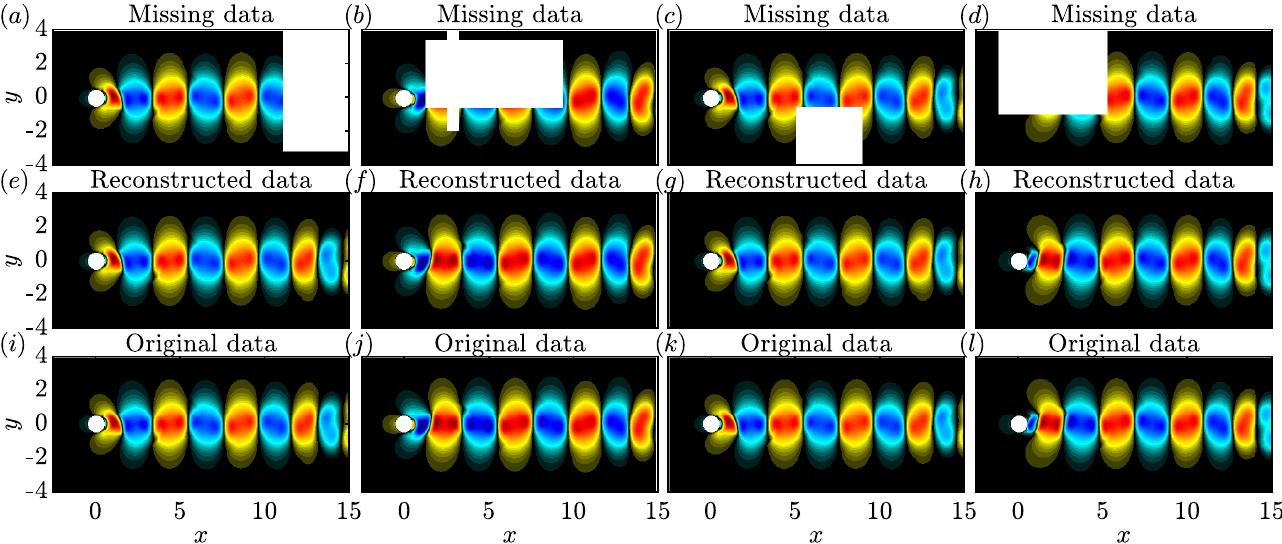}}
\caption{Same as figure \ref{fig_5_per_u_vis} but for the transverse velocity fluctuations.}
\label{fig_5_per_v_vis}
\end{figure}

Figures \ref{fig_5_per_u_vis} and \ref{fig_5_per_v_vis} compare the reconstructed, gappy, and original data for 5\% of missing data. Figure \ref{fig_5_per_u_vis} shows the streamwise velocity fluctuations, $u$, and figure \ref{fig_5_per_v_vis} the transverse component, $v$. The same four time instances are shown in the four columns of both figures. The first column corresponds to the time instant of the largest instantaneous reconstruction error. The other three are randomly selected. Visual inspection of the reconstructed and original data indicates that the gappy SPOD algorithm is able to accurately reconstruct the flow structures in the affected regions. The relative errors for these four snapshots are $e_i = 7\times 10^{-5}$, 1$\times 10^{-5}$, 1$\times 10^{-5}$, and 2$\times 10^{-5}$, respectively. Figures \ref{fig 3.2}-\ref{fig_5_per_v_vis} demonstrate that the algorithm performs very well in both a qualitative and quantitative sense for this case.

\begin{figure}[!th]
\centering
{\includegraphics[trim={0cm 2.25cm 0cm 2.1cm },clip,width=1.0\textwidth]{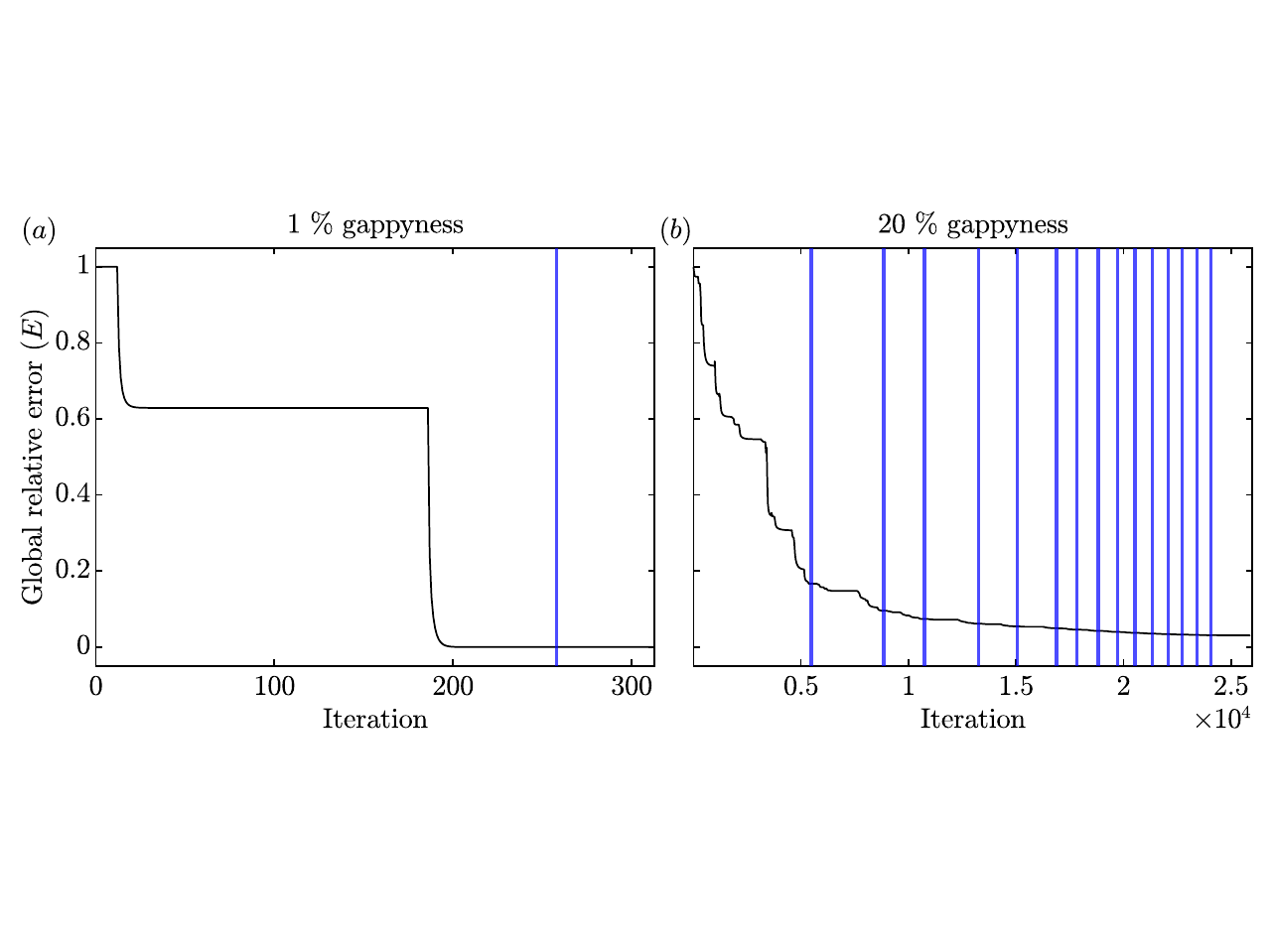}}
\caption{Global relative error of the gappy SPOD algorithm: ($a$) 1\% gappyness; ($b$) 20\% gappyness. The blue solid lines indicate the outer iterations.}
\label{GRE_1_20_cyl}
\end{figure}

We next study the performance of the algorithm for the more moderate case with 1\% and the more extreme case with 20\% of missing data. Figure \ref{GRE_1_20_cyl} shows the global relative errors for these two cases. As in figure \ref{fig 3.2}($a$), outer loops are indicated by thick blue lines. The 1\%- and 20\%-gappyness cases require two and 16 outer loops for convergence, respectively. The final errors upon convergence are $E$ = $9\times10^{-6}$, and $3\times10^{-2}$, respectively. As can be anticipated from the very low error for the 1\% case, the reconstructed velocity fields are visually indistinguishable from the original DNS data, and we hence refrain from showing the reconstructions.

\begin{figure}[h]
\centering
{\includegraphics[trim={0cm 3.45cm 0cm 0.76cm },clip,width=1.0\textwidth]{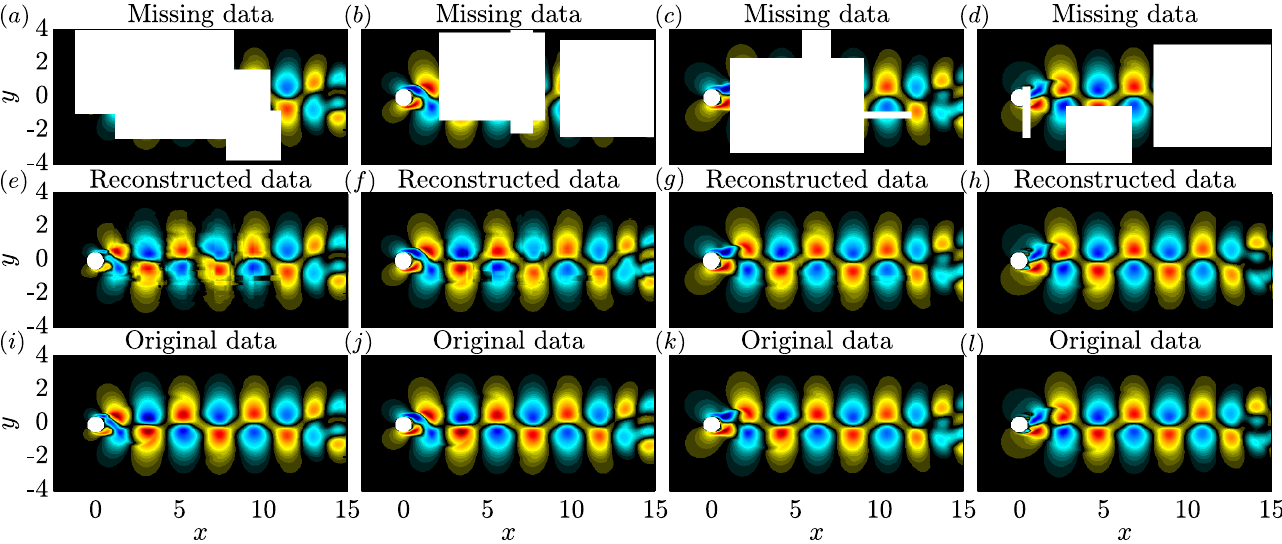}}
\caption{Same as figure \ref{fig_5_per_u_vis}, but for 20\% gappyness.}
\label{fig 3.8}
\end{figure}

Figure \ref{fig 3.8} shows the side-by-side comparison for the most severe case with 20\% of gappyness and a final error of 3\%. While the instantaneous flow fields are shown here, note that the gaps extend over significant time periods, as can be seen in figure \ref{fig 3.1}($c$). Similar to figure \ref{fig_5_per_u_vis}, the first column corresponds to the snapshot that exhibits the largest reconstruction error. The remaining three time instants are arbitrarily selected. The spatial gaps in these four snapshots correspond to 50\%, 46\%, 33\%, and 40\% of missing data, respectively. A comparison of the reconstructed and original fields reveals some local discrepancies at the fringes of the gaps, in particular where multiple gaps overlap. This is most notable in figure \ref{fig 3.8}($e,f$). These local discrepancies aside, the wake flow is accurately reconstructed terms of structure, phase and amplitude. The reconstructions are almost indistinguishable from the original data for the snapshots with 33\% and 40\% of missing data shown in figures \ref{fig 3.8}($c$,$g$,$k$) and \ref{fig 3.8}($d$,$h$,$l$), respectively. Results for the transverse velocity component are very similar and omitted for brevity. The relative errors are $e_i = 7.9\times10^{-2}$, $1.7\times10^{-2}$, $2.5\times10^{-3}$, and $1.3\times10^{-3}$, respectively. Considering that a large fraction of 20\% of the data was missing, these errors are, arguably, low and the overall flow dynamics are accurately recovered. A quantitative comparison with other methods, later presented in \S \ref{Comparison}, confirms this conjecture.

\subsection{Example 2: PIV of turbulent cavity flow by \citet{zhang2017identification,zhang2019spectral,zhang2020spectral}} \label{example 2}

\begin{figure}[h]
\centering
{\includegraphics[trim={0cm 4.5cm 0cm 0.9cm },clip,width=0.9\textwidth]{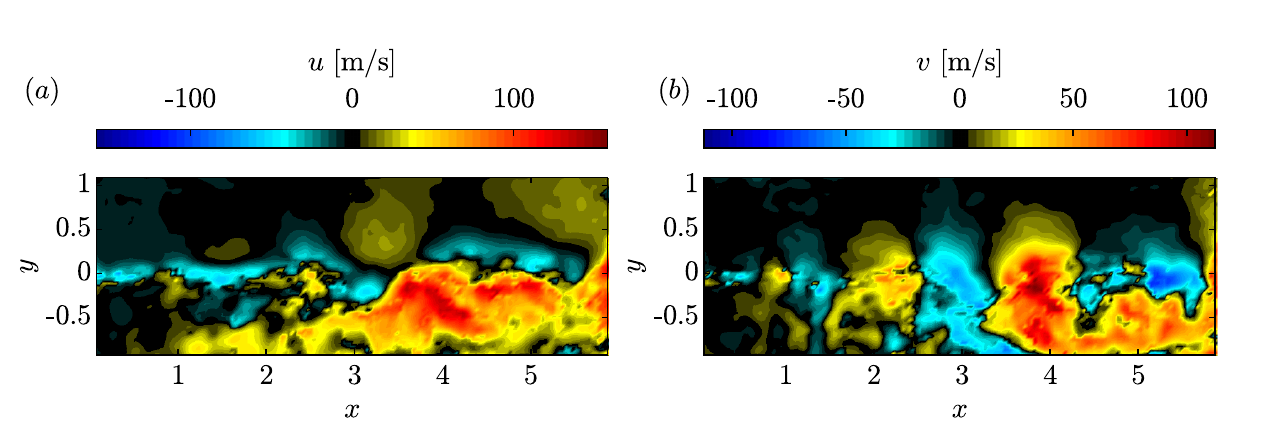}}
\caption{ Instantaneous fluctuating flow field of the turbulent flow over an open cavity at $\Rey$ = 3.3$\times10^5$ measured by \citet{zhang2017identification,zhang2020spectral}: ($a$) streamwise velocity, $u$; ($b$) transverse velocity, $v$. The $x$- and $y$-coordinates are non-dimensionalized by cavity depth $D$.}
\label{fig_Turb_CF}
\end{figure}

Next, we consider the much more relevant and challenging example of experimental turbulent flow over an open cavity. The Reynolds number based on the cavity depth is $\Rey = \rho U_\infty D/\mu=3.3\times10^{5}$, and the Mach number is $M = U_\infty /c_\infty =0.6$. Here, $c_\infty$ is the speed of sound. Time-resolved particle image velocimetry (TR-PIV) was performed to obtain the velocity field in the center plane of an open cavity with a length-to-depth ratio of $L/D = 6$ and a width-to-depth ratio of $W/D = 3.85$. A total number of 16,000 snapshots was obtained at a sampling rate of 16kHz. The coordinates are non-dimensionalized by depth $D$, but the velocity components are reported in SI units. We refer to \citet{zhang2017identification,zhang2019spectral} and \citet{zhang2020spectral} for more details on the measurement campaign and the experimental setup, respectively. The instantaneous velocity field shown in figure \ref{fig_Turb_CF} exemplifies, in stark contrast to the previous example, the chaotic nature of the flow. This example also tests the algorithm's performance in the presence of measurement noise that is not easily distinguished from physical turbulence.

\begin{figure}[h]
\centering
{\includegraphics[trim={0cm 2.8cm 0.5cm 0.45cm },clip,width=1.0\textwidth]{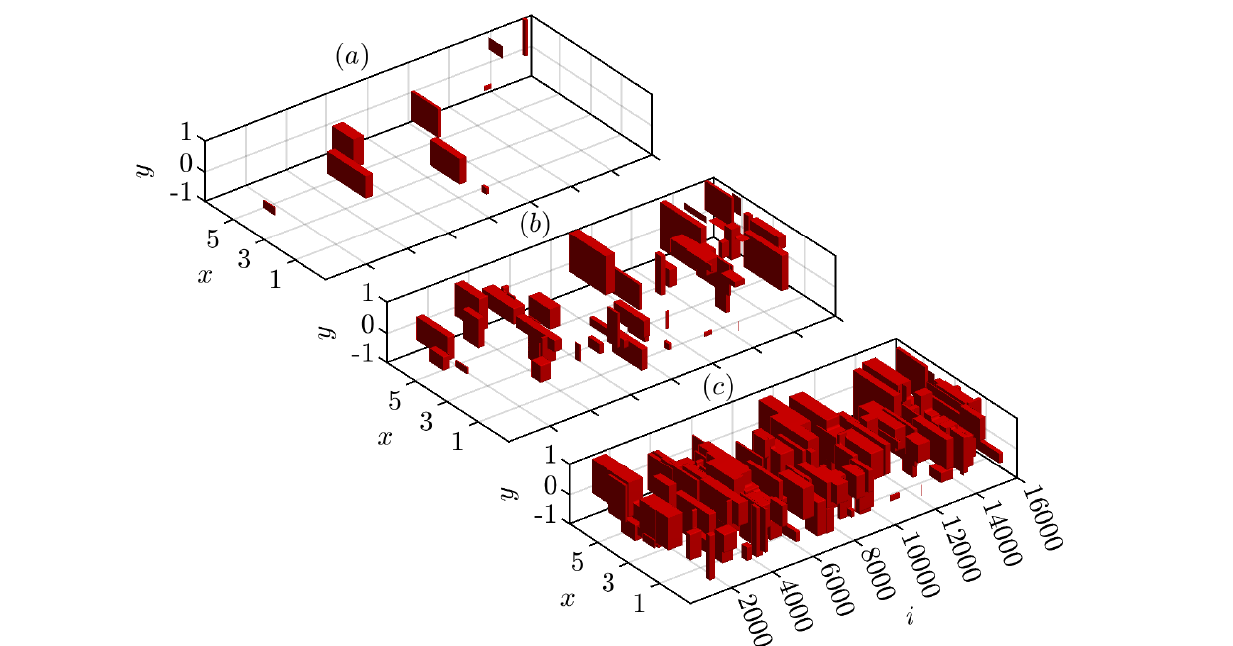}}
\caption{Randomly generated gaps with 5\% for the turbulent flow over an open cavity: ($a$) 1\% gappyness; ($b$) 5\% gappyness; ($c$) 20\% gappyness. Red blocks indicate gaps. The streamwise ($x$-$y$) plane is plotted over the snapshot index, $i$.}
\label{fig_gaps_TC}
\end{figure}

As for the cylinder flow example in \S \ref{example 1}, we consider three cases with 1\%, 5\%, and 20\% of missing data. The gaps are shown in figure \ref{fig_gaps_TC} and were randomly generated in the same way as for the previous example. Taking into account the significantly larger length of the data sequence, the maximum temporal extent of the gaps was increased to 600 snapshots. As before, we choose $n_{\rm fft}=256$ and 50\% overlap, resulting in a total number of 124 blocks. The dependence of the results on $n_{\rm fft}$ is investigated in \ref{Appendix: nfft}. In the extreme example of 20\% gappyness, every block is affected by gaps.  

\begin{figure}[!h]
\centering
{\includegraphics[trim={0cm 1.1cm 0cm 0.08cm },clip,width=1.0\textwidth]{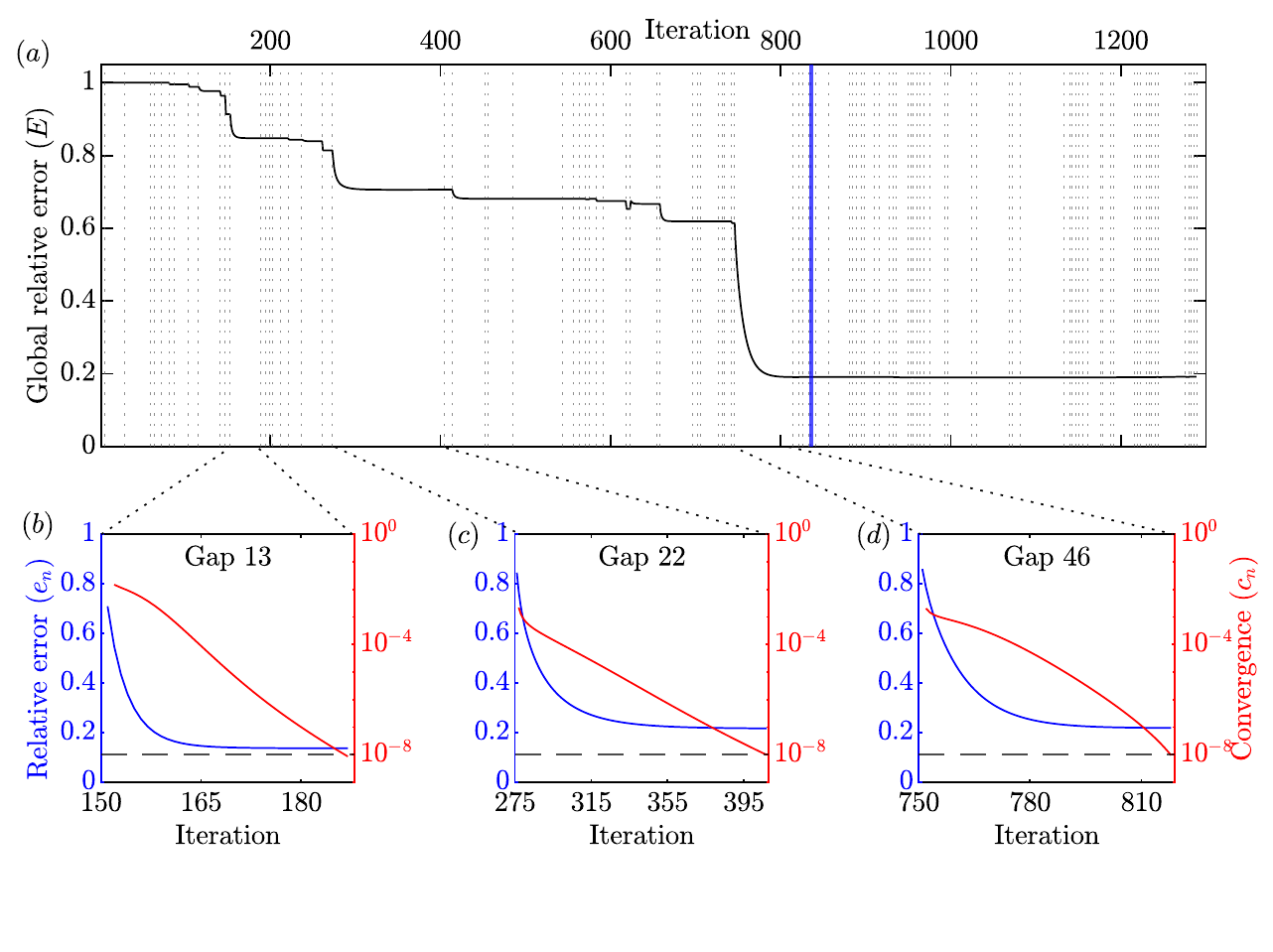}}
\caption{Errors and convergence for 5\% gappyness, see figure \ref{fig_gaps_TC}($b$). The global relative error is shown in ($a$). Inner (gap-wise) iterations are denoted by grey dotted lines and outer iterations as blue solid lines. Panels ($b$-$d$) show the gap-wise error and convergence for three randomly selected gaps within the first outer loop.}
\label{fig_5per_TC}
\end{figure}

As in figure \ref{fig 3.2}, the inner workings of the algorithm for the cavity flow with 5\% of missing data are best understood from the relative errors and gap-wise convergence shown in figure \ref{fig_5per_TC}. Figure \ref{fig_5per_TC}($a$) shows that the algorithm requires two outer iteration loops to satisfy the convergence criterion of $\mathit{tol}=10^{-8}$. The first outer iteration reduces the global relative error by 81\%. The second outer loop does not further reduce the error by an appreciable amount, and its final value remains at $E=19\%$.  In \S \ref{Comparison}, we confirm that this value is lower than what is achieved by other established methods. A notable difference to the laminar case, shown in figure \ref{fig 3.2}, is that the global relative error does not decay monotonically. The local relative error and convergence for three representative gaps during the first outer loop are shown in figure \ref{fig_5per_TC}($b$-$d$). The relative error for all gaps decays to about $20 \%$. An important observation is that the gap-wise relative error always saturates before the convergence criterion of $\mathit{tol}=10^{-8}$ is met. This observation, again, serves as an \emph{a posteriori} justification for this choice of tolerance.

\begin{figure}[!h]
\centering
{\includegraphics[trim={0cm 4.92cm 0cm 0.0cm },clip,width=1.0\textwidth]{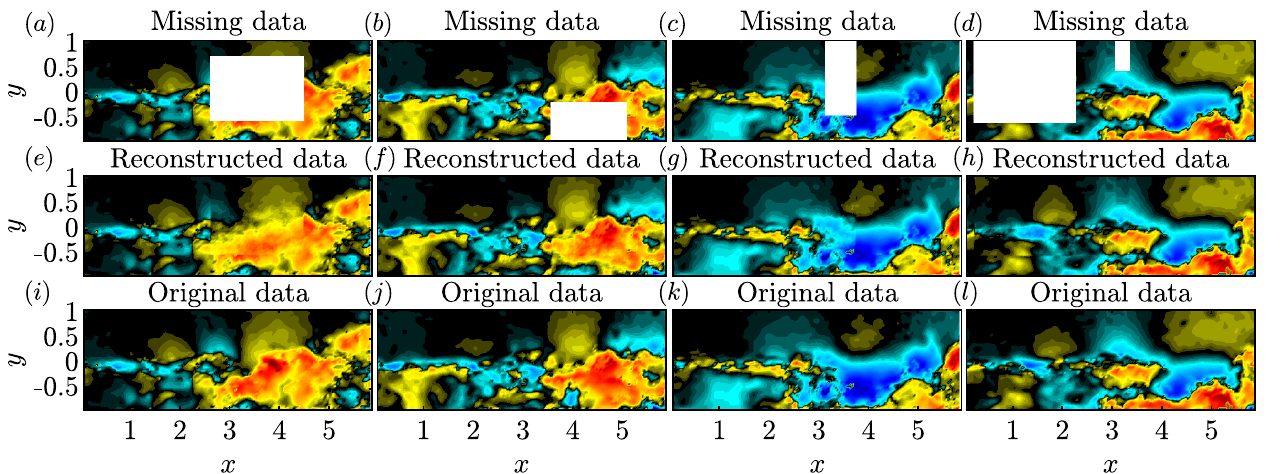}}
\caption{Reconstruction for the turbulent cavity flow with 5\% gappyness at four time instances:  ($a$-$d$) gappy data; ($e$-$h$) reconstructed data; ($i$-$l$) original data. False colors of the streamwise velocity fluctuations, $u$, are on the same scale in all plots.}
\label{fig_5per_vis_TC_u}
\end{figure}

\textbf{\begin{figure}[!h]
\centering
{\includegraphics[trim={0cm 4.92cm 0cm 0.0cm },clip,width=1.0\textwidth]{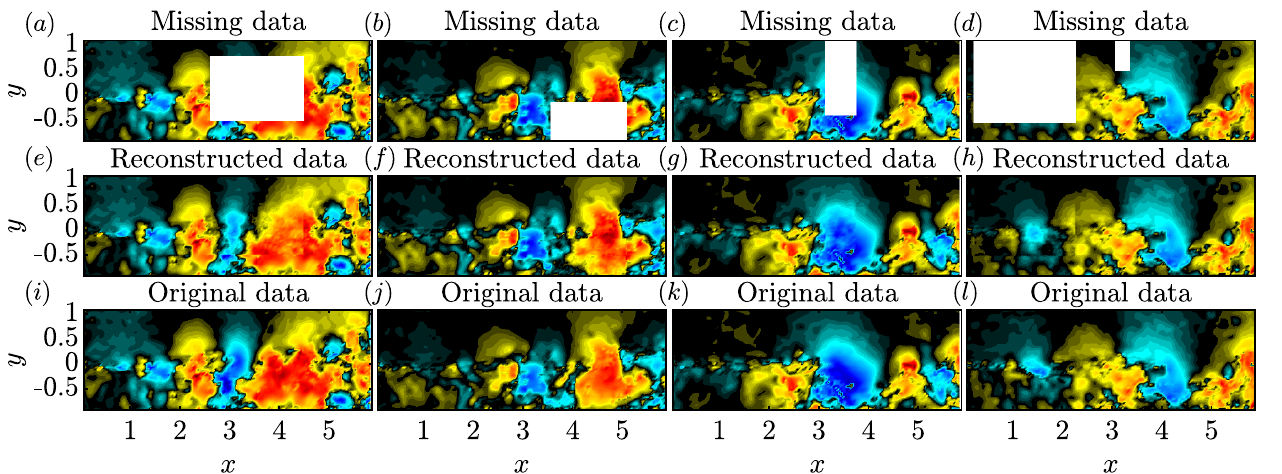}}
\caption{Same as figure \ref{fig_5per_vis_TC_u}, but for the transverse velocity fluctuations.}
\label{fig_5per_vis_TC_v}
\end{figure}}

Figures \ref{fig_5per_vis_TC_u} and \ref{fig_5per_vis_TC_v} show the instantaneous streamwise and transverse velocity components for the gappy, reconstructed and original flow fields. Again, the first column is the snapshot with the largest reconstruction error, and the other three are arbitrarily selected. We observe that the reconstructions are very similar to the original flow field for all four time instances. The corresponding relative errors are $e_i$ = 29\%, 17\%, 8\%, and 22\%, respectively. The comparison of panels ($e$) and ($i$) in figures \ref{fig_5per_vis_TC_u} and \ref{fig_5per_vis_TC_v} shows that the salient flow features are recovered by the algorithm, that is, despite the remaining error of 29\%.

\begin{figure}
\centering
{\includegraphics[trim={0cm 2.2cm 0cm 2.1cm },clip,width=1.0\textwidth]{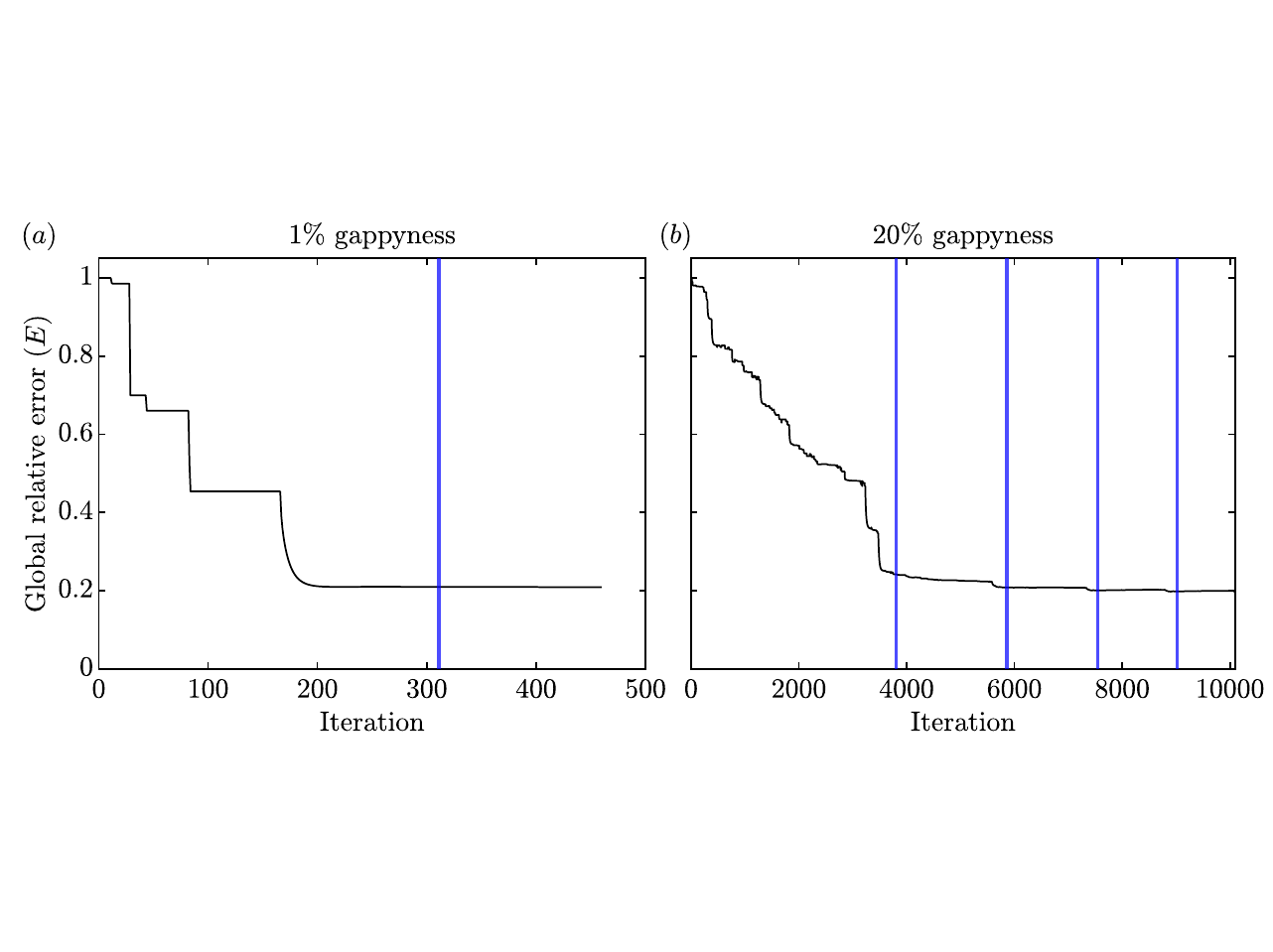}}
\caption{Global relative error of the gappy SPOD algorithm: ($a$) 1\% gappyness; ($b$) 20\% gappyness. The blue solid lines indicate the outer iterations.}
\label{fig_GE_1_20_TC}
\end{figure}

We next address the reconstruction of the two remaining cases with 1\% and 20\% of gappyness. Figure \ref{fig_GE_1_20_TC} shows the global relative errors for these cases. Two and five outer loops are required for convergence, and the final global relative errors are $ E= 20.3\%$ and 19.4\%, respectively. These values are very similar to that of the 5\% case. This indicates that the reconstruction is not sensitive to the percentage of missing data within the range considered here. In each case, the reconstructions recover approximately 80\% of the energy. This result is found in contrast to the laminar cylinder flow case, where the  global relative error proportionally increases with the amount of missing data. As before we next inspect representative time instances for a qualitative assessment of the performance.

\begin{figure}[!h]
\centering
{\includegraphics[trim={0cm 4.9cm 0cm 0.1cm },clip,width=1.0\textwidth]{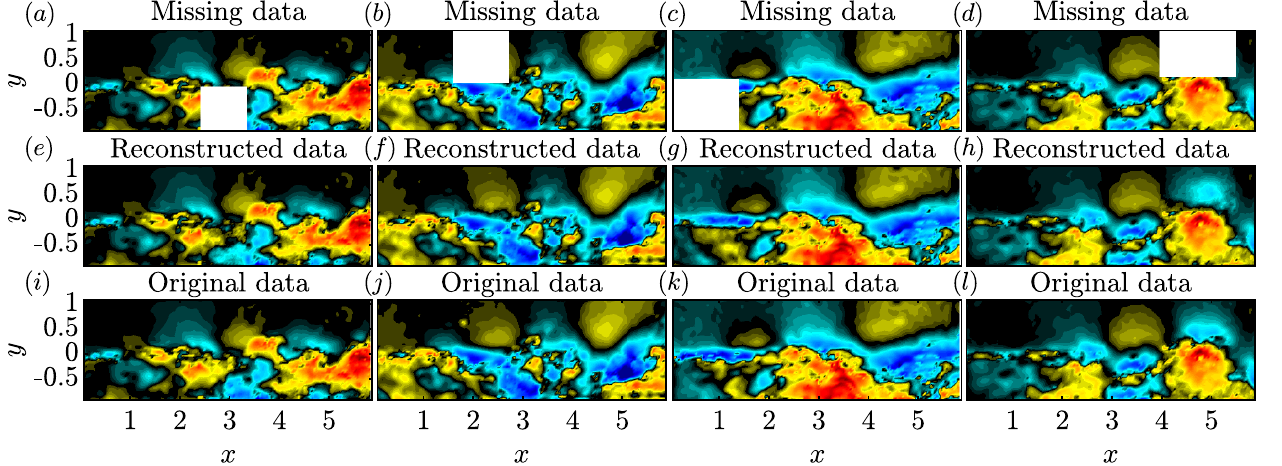}}
\caption{Reconstruction for the turbulent cavity flow with 1\% gappyness at four time instances:  ($a$-$d$) gappy data; ($e$-$h$) reconstructed data; ($i$-$l$) original data. False colors of the streamwise velocity fluctuations, $u$, are on the same scale in all plots.}
\label{fig_1per_TC}
\end{figure}

The reconstructions are compared to the gappy and original flow fields for 1\% and 20\% of missing data in figures \ref{fig_1per_TC} and \ref{fig_20per_TC}, respectively. The reconstructions of the 1\% case are in very good agreement with the original data. For the 20\% case shown in figure \ref{fig_20per_TC}, large parts of the field-of-view are missing. An animation of the snapshots in the vicinity of the gaps (see supplemental material) further confirms that the gaps persist for a long time. The percentages of missing data for the four time instances are 58\%, 48\%, 28\%, and 23\%, and the corresponding relative errors are $e_i$ = 43.2\%, 26.4\%, 13.2\%, and 16.6\%.  A comparison between the reconstructed field-of-view in figure \ref{fig_20per_TC}($e$) and the reconstructed data in \ref{fig_20per_TC}($i$) shows that the gappy SPOD algorithm was able to reconstruct large parts of the flow field. This puts into perspective the reconstruction error of $E=19.4\%$ that might be perceived as large without this qualitative assessment. Note that correlation-based reconstruction is always limited by the physical length and time scales of the turbulent flow beyond which a reconstruction is not feasible. Comparing the reconstructed to the original data for the three remaining time instances confirms that the algorithm is capable of estimating many of the intricate details of the flow.

\textbf{\begin{figure}[!h]
\centering
{\includegraphics[trim={0cm 4.9cm 0cm 0.1cm },clip,width=1.0\textwidth]{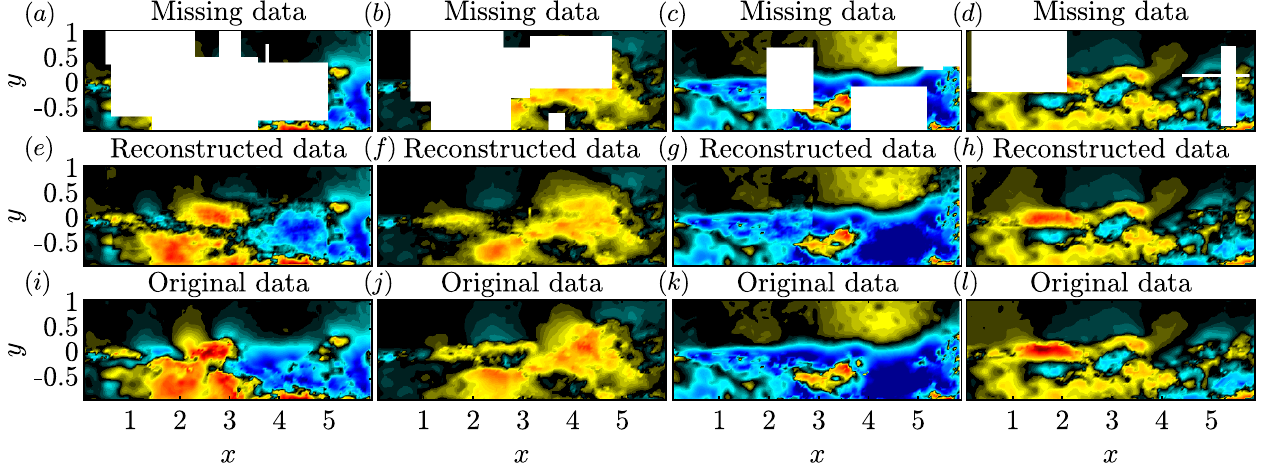}}
\caption{Reconstruction for the turbulent cavity flow with 20\% gappyness at four time instances:  ($a$-$d$) gappy data; ($e$-$h$) reconstructed data; ($i$-$l$) original data. False colors of the streamwise velocity fluctuations, $u$, are on the same scale in all plots.}
\label{fig_20per_TC}
\end{figure}}

\newpage
\subsection{Reconstruction of the turbulence statistics} \label{statistics recon}

\textbf{\begin{figure}[h]
\centering
{\includegraphics[trim={0cm 0cm 0cm 0cm },clip,width=1.0\textwidth]{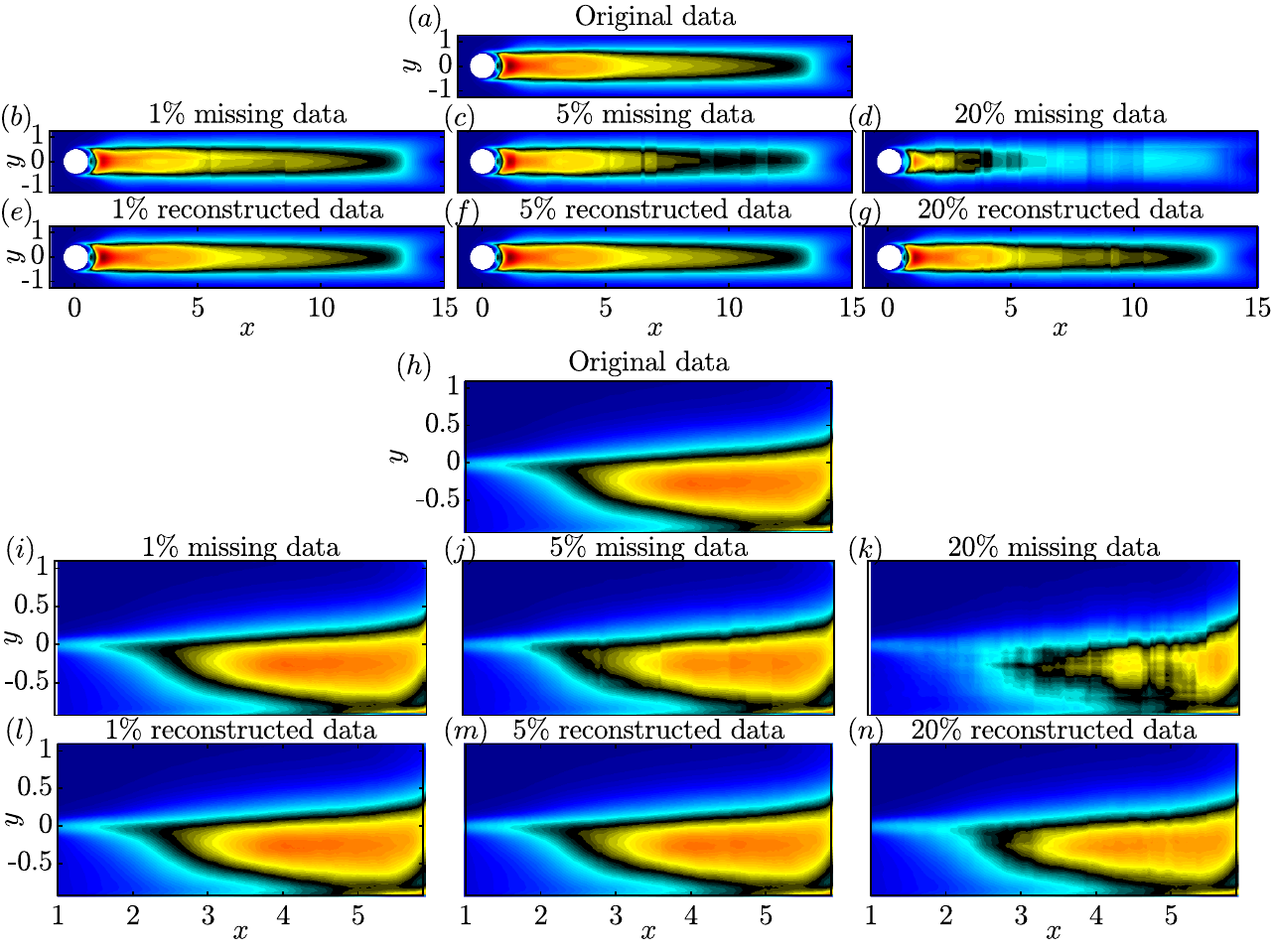}}
\caption{Turbulent kinetic energy fields for the gappy, reconstructed and original data: ($a$-$g$) cylinder flow; ($h$-$n$) the cavity flow.  Contour levels are consistent between subplots ($a$-$g$) and ($h$-$n$).}
\label{fig TKE}
\end{figure}}

After assessing the method's performance in terms of the global reconstruction error and with direct comparisons of original and reconstructed flow fields, we next focus on turbulence quantities. In particular, we consider the turbulent kinetic energy (TKE), TKE = $\frac{1}{2}(\overline{(u^\prime)^2} + \overline{(v^\prime)^2})$, and the Reynolds shear stress, $\tau_{xy} = \overline{u^\prime v^\prime}$. The comparison of the original, gappy, and reconstructed TKE fields for the 1\%, 5\%, and 20\% cases are presented in figure \ref{fig TKE}. The cylinder and cavity flows are shown in figure \ref{fig TKE}($a$-$g$) and \ref{fig TKE}($h$-$n$), respectively. For 1\% and 5\%, moderate distortions are observed in the gappy data for both cases, and the reconstructed TKE fields are almost indistinguishable from the original data. For 20\%, a large fraction of TKE has been removed, and the flow field is heavily distorted. The reconstructions shown in panels ($g$) and ($n$), on the contrary, are smoother, recover large parts of the TKE, and compare well with the original data.

\setlength{\tabcolsep}{5pt}
 \begin{table*}[!h]
\centering
\begin{tabular}{|c|c|c|c|c|c|c|c|c|l} 
\cline{1-9}
\multicolumn{1}{|l|}{}     & \multicolumn{4}{c|}{TKE error (\%)}    & \multicolumn{4}{c|}{Reynolds shear stress error (\%)}   \\ \cline{1-9}
\multirow{2}{*}{Gappyness} & \multicolumn{2}{c|}{Cylinder flow} & \multicolumn{2}{c|}{Cavity flow} &  \multicolumn{2}{c|}{Cylinder flow} & \multicolumn{2}{c|}{Cavity flow}\\ \cline{2-9}
                           & Before   & After           & Before        & After      &  Before   & After     &  Before   & After      &\\ \cline{1-9}
1\%                        & 1.18\%    & 3.40$\times 10^{-3}$ \% & 0.65\%         & 0.19\%     & 1.67\%    & 2.68 $\times 10^{-2}$\% & 0.65\%       & 0.14\%&  \\ \cline{1-9}
5\%                        & 8.23\%     & 3.49$\times 10^{-2}$ \% & 4.05\%         & 1.3\%      & 5.74\%     & 0.18\%   & 4.27\%         & 1.02\%&  \\ \cline{1-9}
20\%                       & 34.72\%   & 4.51\%                  & 21.96\%        & 6.35\%      & 25.25\%   & 5.92\%                & 24.59\%        & 5.56\%  &  \\  \cline{1-9} 
\end{tabular}
\caption{\label{table TKE and Rey error} Area integrated TKE and Reynolds stress errors for the gappy and reconstructed flow fields.}
\end{table*}   

For a more quantitative assessment, the area-integrated TKE error is presented in table \ref{table TKE and Rey error}. In accordance with figure \ref{fig TKE}, the TKE error after reconstruction is almost negligible for 1\% and 5\% of missing data for the cylinder flow. The TKE error is also substantially lower for the cavity flow. For 20\% of missing data, the TKE error reduces from 34.7\% to 4.5\%, and from 22\% to 6.4\% for the two examples, respectively. These percentages correspond to $\approx7$- and 4-fold reductions. The qualitative counterparts to these numbers are the TKE field comparisons previously shown in figure \ref{fig TKE}($d,g$) and \ref{fig TKE}($k,n$), respectively. Also listed in table \ref{table TKE and Rey error} is the area-integrated Reynolds shear stress error. Similar to the TKE error, the Reynolds shear stress error has been reduced significantly by a factor of $\approx4$ in both cases.

\section{Comparison with other methods} \label{Comparison}
\begin{figure}
\centering
{\includegraphics[trim={0cm 1.45cm 0cm 0.8cm },clip,width=1.0\textwidth]{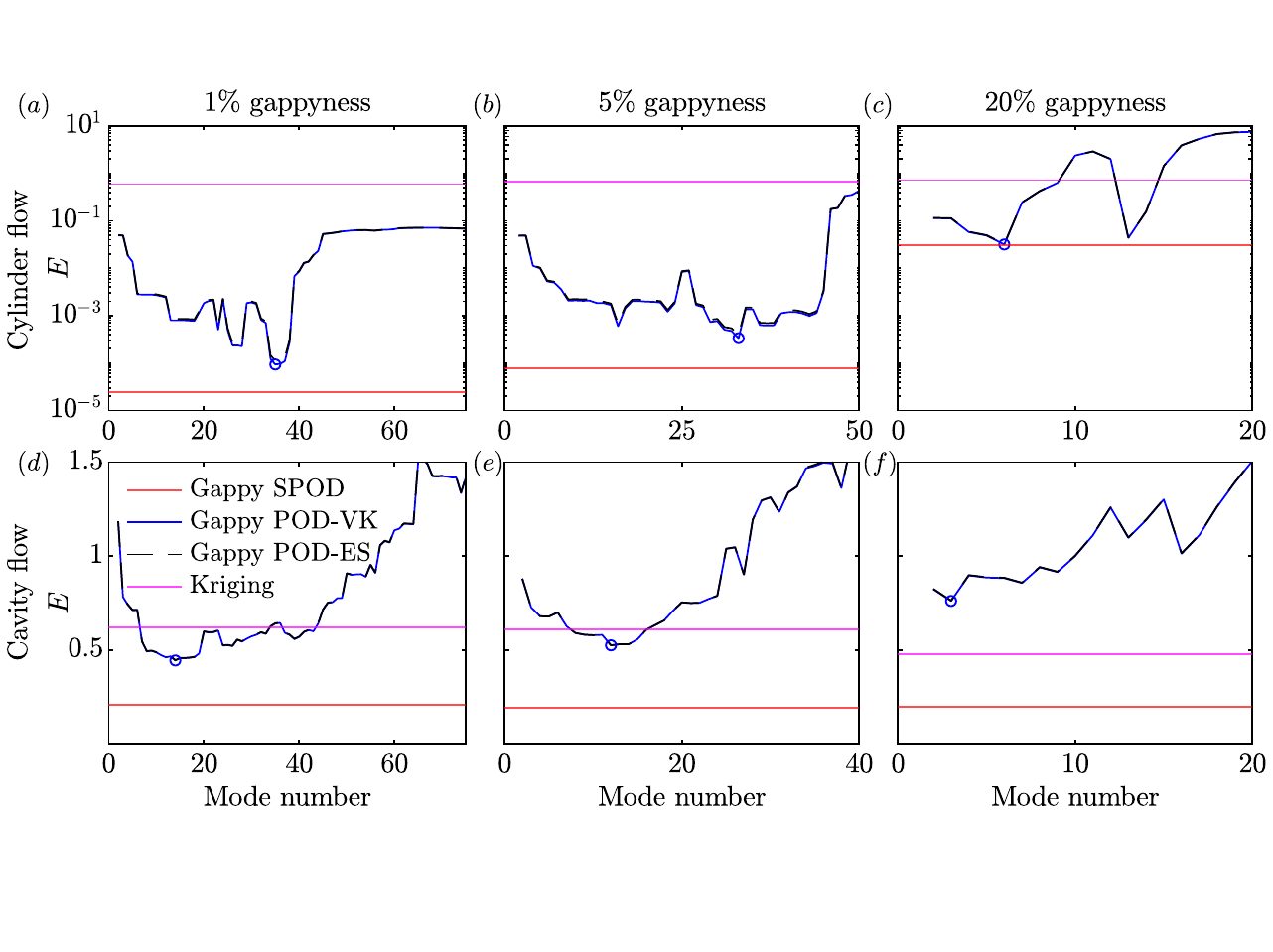}}\caption{Global relative error for different reconstruction methods: ($a$-$c$) cylinder flow; ($d$-$f$) cavity flow; ($a$,$d$) 1\%, ($b$,$e$) 5\% and ($c$,$f$) 20\% of missing data. The ordinate in ($a$-$c$)is on a logarithmic scale for clarity.}
\label{fig methods comparison}
\end{figure}

Finally, we compare the performance of the gappy SPOD method to the established methods of gappy POD by \citet{everson1995karhunen} and \citet{venturi2004gappy} (referred to as gappy POD-ES and -VK in the following), and Kriging. For local Kriging, the exponential correlation model results, in agreement with \citet{saini2016development}, in lower errors than the more standard Gaussian model. The more successful model is implemented in the MATLAB toolbox DACE \citep{lophaven2002dace} that was used for our comparisons.  The same tool was used by \citet{venturi2004gappy}, \citet{raben2012adaptive}, and \citet{saini2016development} for the same purpose.

Figure \ref{fig methods comparison} compares the global relative errors for gappy SPOD, the two gappy POD algorithms, and Kriging. Both cases and all three levels of gappyness are considered. Both gappy POD methods have the number of modes as a free parameter that needs to be varied to identify the best-possible reconstruction. Kriging relies on interpolants instead of modes and gappy SPOD does not truncate the modal basis. The reconstruction errors of these methods are therefore constant in figure \ref{fig methods comparison}. The gappy POD-ES and -VK methods produce almost identical errors \citep[see also][]{raben2012adaptive,saini2016development}, and we hence do not distinguish between them. The mode numbers were varied until candidates for the global minima, marked by circles, were identified as minima by observing the general trend of the errors. As the number of modes in classical POD is equal to the number of snapshots (4096 and 16000 for the data at hand), gappy POD becomes computationally intractable due to increasingly high compute times (see table 3 below). Therefore, there is no guarantee that the reconstruction corresponds to the actual global optimum. A few common trends are observed in all six cases in figure \ref{fig methods comparison}. In general, gappy POD outperforms Kriging and gappy SPOD outperforms gappy POD. For the cylinder flow and 1\% and 5\% of gappyness in figure \ref{fig methods comparison}($a$,$b$), final reconstruction errors for gappy SPOD are 3.8 and 4.3 times below those of gappy POD and multiple orders of magnitude lower than those of Kriging. For 20\% gappyness, gappy SPOD and gappy POD yield very similar results with only a marginal advantage of 2\% for gappy SPOD, but both outperform Kriging by about one order of magnitude. For the turbulent cavity flow in figure \ref{fig methods comparison}($d$-$f$), at all missing data percentages, gappy SPOD reduces the global relative error by about 80\%, which is at least 20\% lower than for other methods in all cases. In direct comparison with gappy POD, gappy SPOD achieves 2.1, 2.7, and 3.8 fold reductions in error for the three gappyness ratios, respectively. As in section \S\S \ref{example 1} and \ref{example 2}, we observe that the gappyness has a significant impact on the achievable error reduction for the periodic laminar flow. In contrast, the final error is almost independent of the percentage gappyness for the turbulent, noisy experimental data.

We next compare the compute times for Kriging, gappy POD, and gappy SPOD in table \ref{table time Taken}.  With the exception of the cylinder flow at the two lower percentages (where gappy SPOD has an advantage), Kriging and gappy SPOD have similar computational times. A direct comparison with gappy POD is not possible as the optimal number of modes is not known a priori, and its computation becomes quickly intractable as the mode number is increased.  The mode number shown in table \ref{table time Taken} reflect the minimum number of modes required to guarantee that the global minimum was identified to the best of our ability. For the two examples at hand, we found that a decreasing number of modes was required for gappy POD as the gappyness increases. This gives gappy POD an advantage in terms of compute time for higher percentages of missing data. Note, however, that gappy SPOD yields a more accurate reconstruction, and does not require monitoring. In summary, gappy SPOD always outperforms Kriging and generally performs better (although in one case only marginally) than gappy POD in terms of the reconstruction error, but sometimes does so at a higher computational cost.
    

\setlength{\tabcolsep}{5pt}
\begin{table*}
\centering
\begin{tabular}{|c|c|c|c|c|c|c|l} 
\cline{1-7}
{Gappy-}     & \multicolumn{3}{c|}{Cylinder flow (in hr)}    & \multicolumn{3}{c|}{Cavity flow (in hr)}   \\ \cline{2-7}
{ness} & Kriging & gappy POD & gappy SPOD &  Kriging & gappy POD & gappy SPOD \\ \cline{1-7}
                         
1\%                        &  11.24    & 49.95 (75 modes) & 1.63           &   2.79    & 27.88 (75 modes)    & 2.37   & \\ \cline{1-7}
5\%                        &  28.91      &  64.69 (50 modes) & 13.58           &    10.91     & 19.94 (40 modes)  & 7.28     &  \\ \cline{1-7}
20\%                       &   159.63   &    34.13 (20 modes)               & 139.86         & 46.84        & 26.30 (20 modes)   &  49.57                 &   \\  \cline{1-7} 
\end{tabular}
\caption{\label{table time Taken} Comparison of computational time. The computations were performed on a high-performance workstation with 192GB of memory and two 3.0 GHz 48-core Intel Xeon Gold CPUs. The spatial degrees of freedom times the number of variables for the cylinder and cavity flows are 62500, and 16500, respectively. The corresponding number of snapshots are 4096 and 16000.}
\end{table*}

\section{Summary and discussion} \label{conclusion}

A new algorithm is proposed that leverages the temporal correlation of SPOD modes with preceding and succeeding snapshots and their spatial correlation with the surrounding data to reconstruct partially missing or corrupted flow data. For demonstration purposes only, the reconstructed data are compared to the actual data in the corrupted regions. The algorithm itself exclusively relies on convergence metrics and fills in the gaps sequentially until the reconstruction converges to a user-defined tolerance, both locally, that is for each gap, and globally. The method is demonstrated on simulation data of the flow around a cylinder and time-resolved PIV data of turbulent cavity flow; the first being a canonical benchmark example used in many previous studies and the latter a realistic scenario of turbulent flow data in the presence of measurement noise. For randomly seeded and sized gaps that amount to up to 20\% of missing data and extend over large regions in space and many snapshots, the algorithm accurately recovers the missing instantaneous ans mean flow fields as well as turbulence statistics. It generally outperforms the established methods, particularly for the turbulent flow, where it yields a significantly lower reconstruction error that translates into an at least two-fold reduction compared to gappy POD and Kriging. Notably, this comparably low reconstruction error appears largely unaffected by the percentage gappyness within the range tested here. Even though the different methods scale differently and the results are case-dependent, their computational costs are roughly comparable. A caveat to the strong performance of the new method is that it is strictly applicable to statistically stationary data only, and that it relies on a sufficiently well-converged SPOD of the data, which in turn requires a sufficiently long time series. The main limitation of the algorithm directly follows from the properties of the SPOD: accurate flow reconstruction is only possible within the physical correlation length and time scales of the flow. This limitation, however, applies to all correlation-based methods. A systematic study of the most important spectral estimation parameters shows that best practices for SPOD \citep{schmidt2020guide} also lead to a good balance between the accuracy and computational cost of the gappy SPOD algorithm. In \ref{Appendix: noise_added}, we demonstrate that the algorithm not only performs well in the presence of noise, but also that mode truncation facilitates efficient denoising while reducing computational time. Contingent on further testing on more data sets, the present results suggest that the algorithm can be fully automated and applied to any sufficiently long stationary flow data. Future extensions of gappy SPOD may include algorithmic improvements for specific error types like missing snapshots and the inclusion of trends for transient, non-stationary data.


\textbf{Acknowledgements} We gratefully acknowledge partial support from Office of Naval Research grant N00014-20-1-2311 and AFOSR Grant FA9550-22-1-0541. We thank Lou Cattafesta and Yang Zhang for providing the TR-PIV data and fruitful discussions. The data was created with supported by the U.S. Air Force Office of Scientific Research award FA9550-17-1-0380.

\appendix
\section{Missing snapshots} \label{Appendix: miss snapshots}

\begin{figure}
\centering
{\includegraphics[trim={0cm 2.25cm 0cm 2.2cm },clip,width=1.0\textwidth]{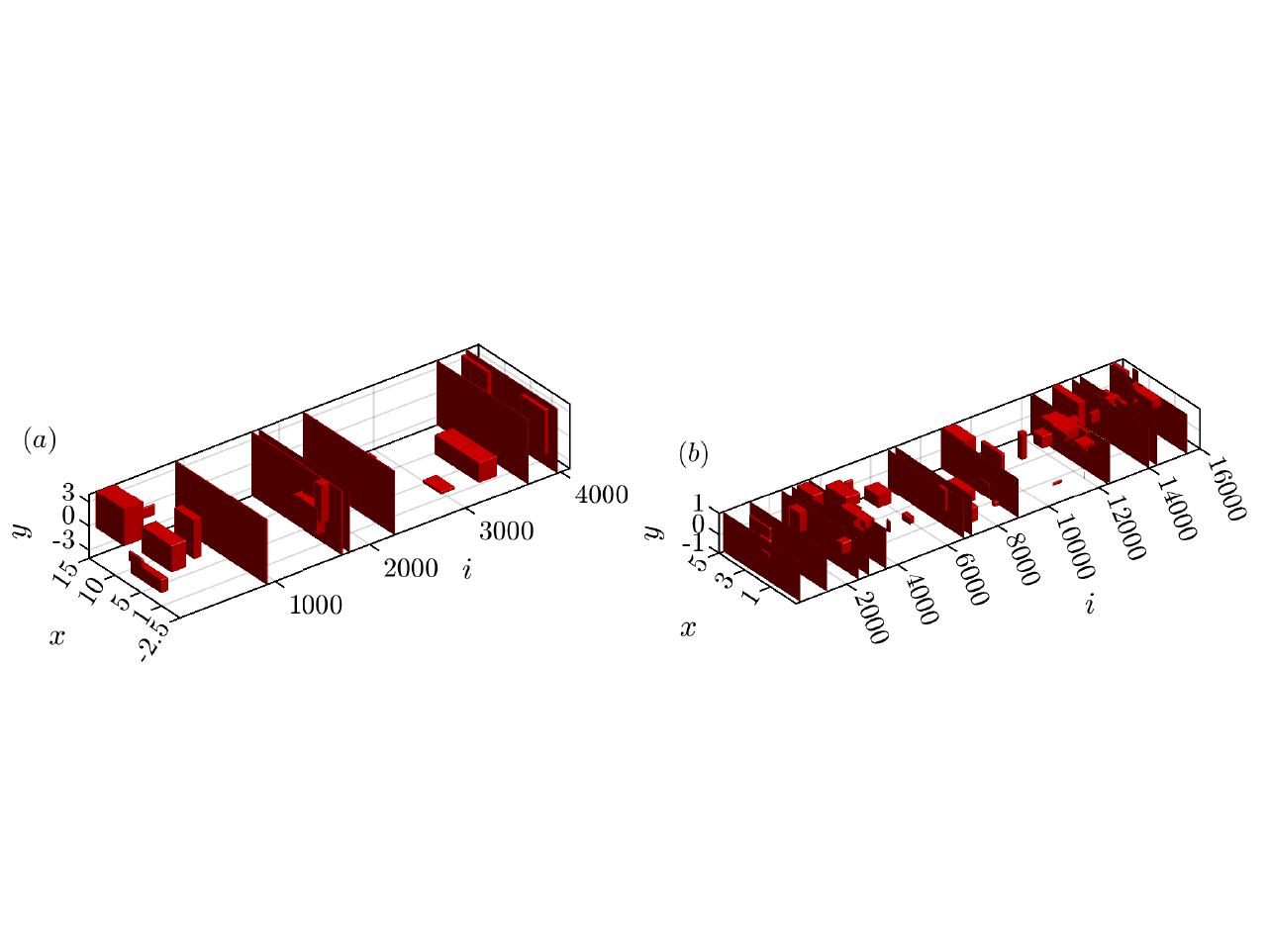}}
\caption{Randomly generated gaps with 5\% of missing data and missing snapshots: ($a$) cylinder flow and ($b$) cavity flow.}
\label{fig miss snapshots}
\end{figure}

A common scenario not considered so far is that of missing snapshots, or even longer sequences of missing snapshots. Unlike the classical gappy POD methods \citep{everson1995karhunen,venturi2004gappy}, the gappy SPOD algorithm does not rely on a least-squares fit and hence does not become singular in this situation. Figure \ref{fig miss snapshots} shows the combination of gaps and missing snapshots investigated here, and which amounts to 5\% of missing data. The gaps are randomly generated in the same way as in \S\S \ref{example 1} and \ref{example 2}. For isolated, non-consecutive missing snapshots, temporal interpolation can provide accurate reconstructions, and we hence do not consider this simpler scenario. Instead, we add for the cylinder and cavity flow, respectively, five and ten sequences of ten to 25 consecutive missing snapshots. The start of the missing snapshot sequences and their duration are randomly selected within this interval. The main difference between localized gaps and missing snapshot sequences is that, in the latter, the algorithm solely relies on temporal correlation with previous and subsequent snapshots for the reconstruction.

\begin{figure}
\centering
{\includegraphics[trim={0cm 2.25cm 0cm 2.2cm },clip,width=1.0\textwidth]{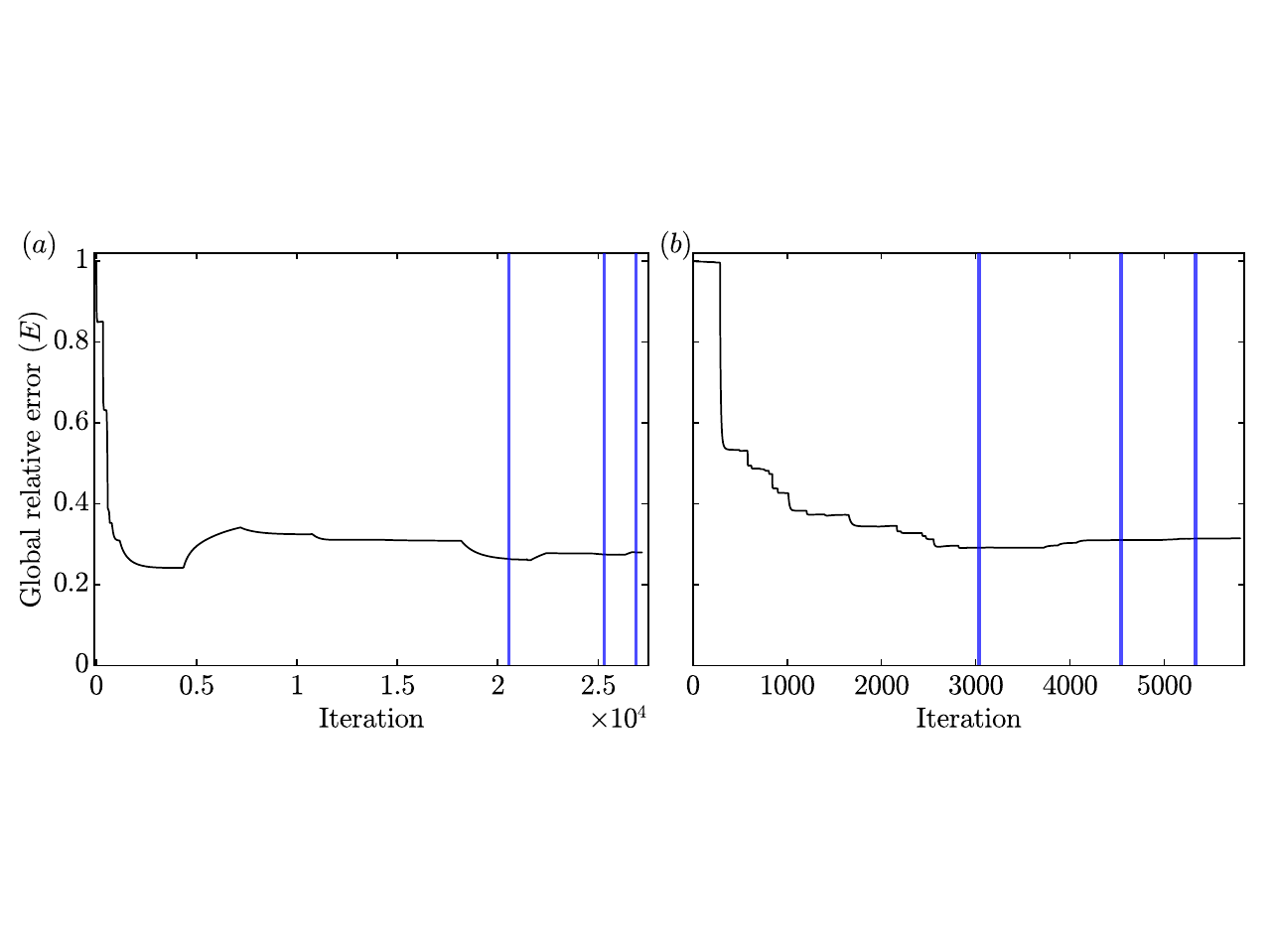}}
\caption{Global relative error for cases with missing snapshots shown in figure \ref{fig miss snapshots}: ($a$) cylinder flow; ($b$) cavity flow. }
\label{fig error miss snapshots}
\end{figure}

The global relative errors for this scenario are shown in figure \ref{fig error miss snapshots}. As before, the tolerance is set to $\mathit{tol}=10^{-8}$ and both cases require four outer loops for convergence. The final errors are $E=28\%$, and 31\%, respectively. For the cavity flow example, the final error, although higher, is still comparable to that of the 5\% case shown in figure \ref{fig_5per_TC}. For the cylinder flow example, on the other hand, the final error is comparable to that of the turbulent cavity flow, but significantly higher than the corresponding case shown in figure \ref{fig 3.2}. A notable difference compared to cases without missing snapshots is that the converged solution does not necessarily yield the minimum global relative error. However, the converged values are within 4\% of the global minimum, and about 70\% of the missing energy is recovered for both examples. Missing snapshots apparently pose a special challenge to the algorithm, in particular for the laminar cylinder flow. An inspection of the reconstructed data before and after the largest jump occurring at iteration 4340 in figure \ref{fig error miss snapshots}($a$)  reveals that the jump is caused by an inversion of the phase of the entire flow field, and that the jump occurs for the gap with the longest consecutive sequence of 25 missing snapshots. For simplicity, we refrain from introducing a special treatment of this very specific phase-inversion error and emphasize that the final error is comparable to its global minimum. A special treatment of this error type is a direction of future research.

\section{Effect of spectral estimation parameters $n_{\rm{fft}}$ and $n_{\rm{ovlp}}$ } \label{Appendix: nfft}
\begin{figure}[h]
\centering
{\includegraphics[trim={0cm 1.0cm 0cm 0.90cm },clip,width=1.0\textwidth]{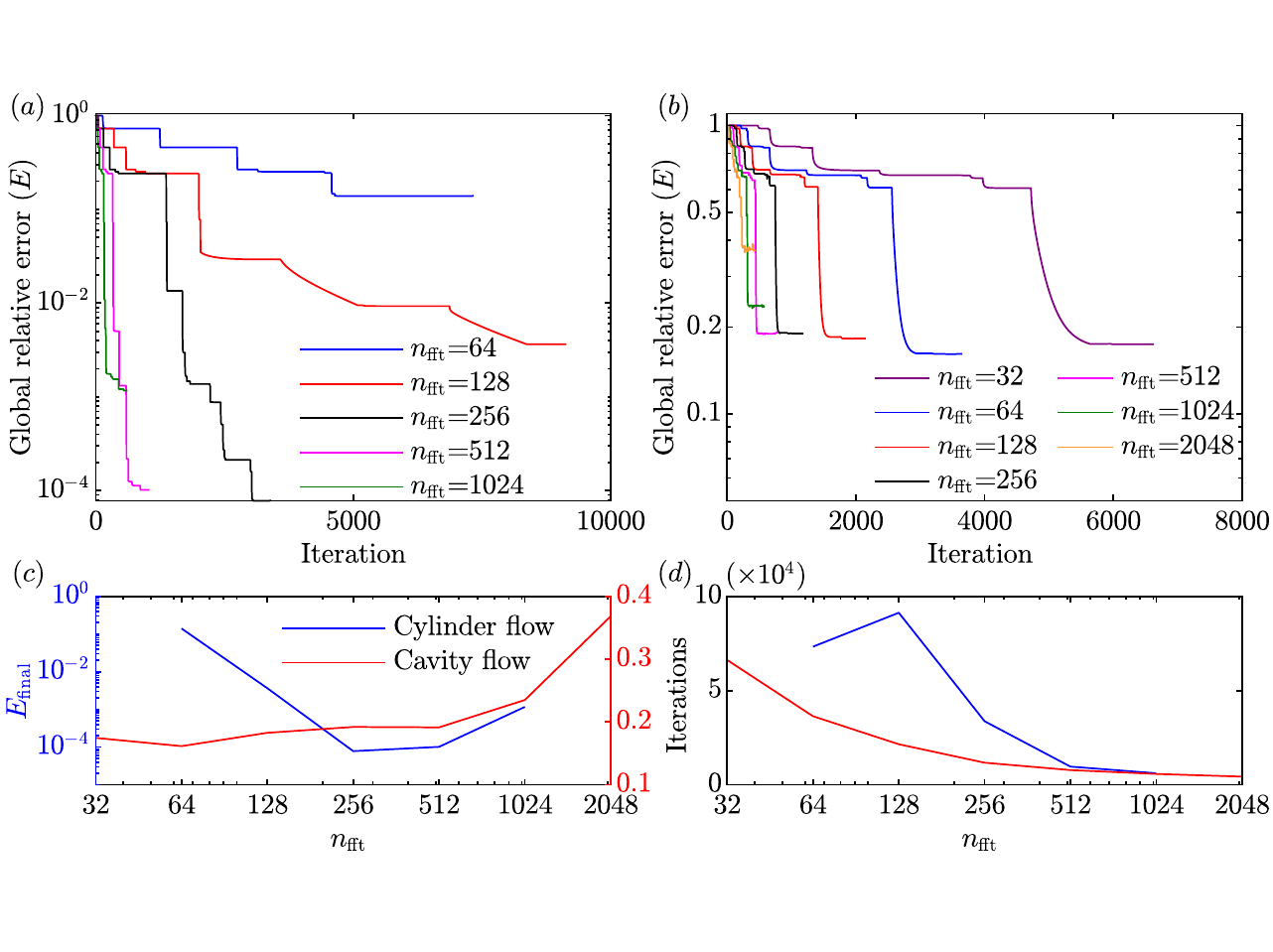}}
\caption{Effect of $n_{\rm{fft}}$ on the global relative error: ($a$) cylinder flow; ($b$) cavity flow; ($c$) final error, and ($d$), total number of iterations as a function of $n_{\rm{fft}}$.  }
\label{fig nfft study}
\end{figure}

The arguably most crucial aspect of spectral estimation is the inevitable trade-off between variance and bias. In SPOD, this trade-off is mainly determined by the block size, $n_{\rm{fft}}$.  For low $n_\textrm{fft}$, the frequency bin size, $\Delta f \propto 1/n_\textrm{fft}$, is large and the spectrum will be biased. For large $n_\textrm{fft}$, the total number of blocks, and therefore realizations of the Fourier transform, is low and the spectrum suffers from high variance. The number of blocks can be inflated by increasing the overlap, $n_{\rm{ovlp}}$. However, values above $50\%$ in practice do not yield lower-variance estimates as the information contained in neighbouring blocks becomes increasingly redundant \citep{schmidt2020guide}. Figure \ref{fig nfft study} summarizes a parameter study of $n_\textrm{fft}$ for both test cases.  As in figures \ref{fig 3.2} and \ref{fig_5per_TC}, we show the global relative reconstruction errors for varying $n_\textrm{fft}$ and 5\% gappyness. Results for the cylinder and cavity flow cases are shown in figure \ref{fig nfft study}($a$) and \ref{fig nfft study}($b$), respectively. The final error is more sensitive for the laminar flow and is hence reported on a logarithmic scale. As before, all computations were converged to a tolerance of $\mathit{tol}=10^{-8}$. The final errors, $E_{\rm{final}}$,  and total number of iterations, are extracted and reproduced in figure \ref{fig nfft study}($c$) and \ref{fig nfft study}($d$). In both cases, the global relative error first decreases and then increases with $n_\textrm{fft}$. The lowest errors are achieved for $n_\textrm{fft}=256$, and $n_\textrm{fft}=64$, respectively. With the exception of $n_\textrm{fft} =64$ for the cylinder flow, the number of iterations decreases monotonically with $n_\textrm{fft}$. This can be seen in figure \ref{fig nfft study}($d$). The number of iterations, in turn, is directly indicative of the overall compute time. In the main text, we did not use this \emph{a posteriori} knowledge of the optimal value of $n_\textrm{fft}$. For the cylinder, the best practise value of $n_\textrm{fft}=256$ yields the minimum error. For the cavity flow this is not the case. However, the error for $32 \leq n_\textrm{fft} \leq 512$ only deviates within $3\%$ from the optimal value. At the same time, using $n_\textrm{fft}=256$ leads to a factor 3.5 savings in compute time as compared $n_\textrm{fft}=64$. To summarize, the best practise of $n_\textrm{fft}=256$ yielded the optimal value for one of the examples, and a very good compromise between reconstruction accuracy and compute time in the other.

\begin{figure}[h]
\centering
{\includegraphics[trim={0cm 1.0cm 0cm 5.1cm },clip,width=1.0\textwidth]{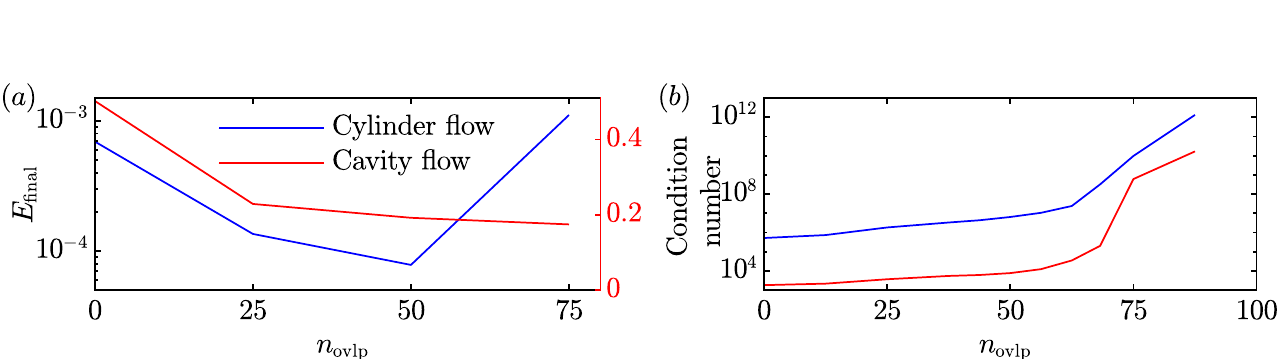}}
\caption{Effect of $n_{\rm{ovlp}}$ on the, ($a$), global relative error and, ($b$), condition number of the weighted CSD matrices for 5\% missing data. The maximum condition number over all frequencies is shown.}
\label{fig novlp study}
\end{figure}

 The effect of the overlap, $n_{\rm{ovlp}}$, is investigated in figure \ref{fig novlp study} for a fixed segment length of $n_{\rm{fft}}=256$ and 5\% gappyness. The final reconstruction errors for $n_{\rm{ovlp}}=0\%$, 25\%, 50\% and 75\% are shown in figure \ref{fig novlp study}($a$). For the turbulent cavity flow, the error decreases with increasing overlap. For the laminar cylinder wake, the error first decreases, but then increases again for the largest overlap of $n_{\rm{ovlp}}=75\%$. Notably, the best-practice value of $n_{\rm{ovlp}}=50\%$ yields the lowest error for the cylinder flow and is within 2\% of the optimum achieved for the cavity flow. The latter is achieved for $n_{\rm{ovlp}}=75\%$ and comes with an approximately 50\% increase in computational cost. The increase of the error at $n_{\rm{ovlp}}=75\%$ for the laminar cylinder wake is unexpected at first. We speculate that it is related to the increasing redundancy of information in the overlapping blocks, which in turn results in increasing linear dependence. The latter can be gauged by the condition number. In figure \ref{fig novlp study}($b$), the maximum condition number over all frequencies, $\max_l(\lambda_l^{(1)}/\lambda_l^{(n_{\rm{blk}})})$, is shown as a function of overlap. As the condition number is independent of the data reconstruction, more data points are computed and shown in figure \ref{fig novlp study}($b$). The trends for the two flows are similar, but the condition number of the cylinder flows is, indeed, significantly higher and elevated by about two orders of magnitude for $n_{\rm{ovlp}} \lesssim 60\%$. For larger values, the condition numbers of both flows suddenly increase. These observations support the best-practice value of $n_{\rm{ovlp}}= 50\%$, which stays well clear of ill-conditioning associated with high overlaps.

\section{Effect of data length $n_{t}$ } \label{Appendix: convergence}
\begin{figure}
\centering
{\includegraphics[trim={0cm 1.42cm 0cm 3.8cm },clip,width=1.0\textwidth]{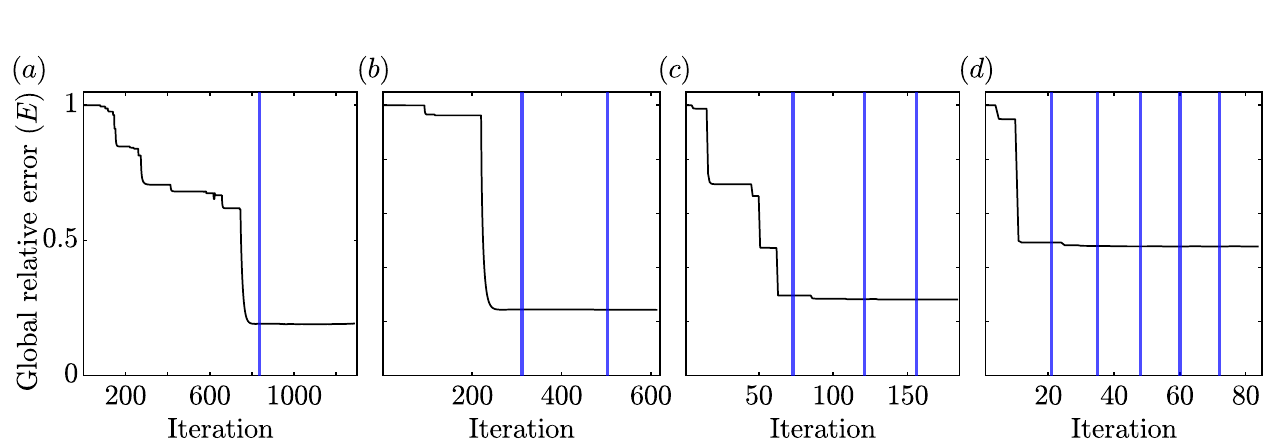}}
\caption{Global relative error for the cavity flow with 5\% of missing data: ($a$) full data; ($b$) first half of the data;  ($c$) first quarter of the data; ($d$) first 10\% of the data. The total number of snapshots are 16000, 8000, 4000, and 1600, respectively.  }
\label{fig SPOD convergence}
\end{figure}

As a statistical method, SPOD relies on a large ensemble realizations of the Fourier transform to converge the modes and eigenvalues. We hence expect the quality of the reconstruction to be better for longer time series. The following study of increasingly truncated subsets of the 5\% gappyness cavity data confirms this. The original example shown in figure \ref{fig_gaps_TC}($b$) is truncated by 50\%, 75\%, and 90\%.  The convergence in terms of the global relative errors for the full and truncated datasets are shown in figure \ref{fig SPOD convergence}. For comparability, $n_{\rm{fft}}$ is fixed at 256. Upon convergence, the final global relative error increases with increasing truncation from $E=19\%$ to 24.3\%, 28.1\%, and finally 47.8\%, as anticipated. At the same time, we observe that the shorter datasets require a higher number of outer iterations for global convergence. The full and truncated datasets require two, three, four, and six outer iterations, respectively. These observations confirm the dependence of gappy SPOD algorithm on a sufficiently long time series and, therefore, sufficiently converged SPOD modes. As a rule of thumb, this suggests that gappy SPOD should only be applied to datasets that are suitable for SPOD analysis in the first place.

\section{Performance in the presence of noise and filtering}  \label{Appendix: noise_added}
\begin{figure}[h!]
\centering
{\includegraphics[trim={0cm 0.1cm 0cm 0.6cm },clip,width=1.0\textwidth]{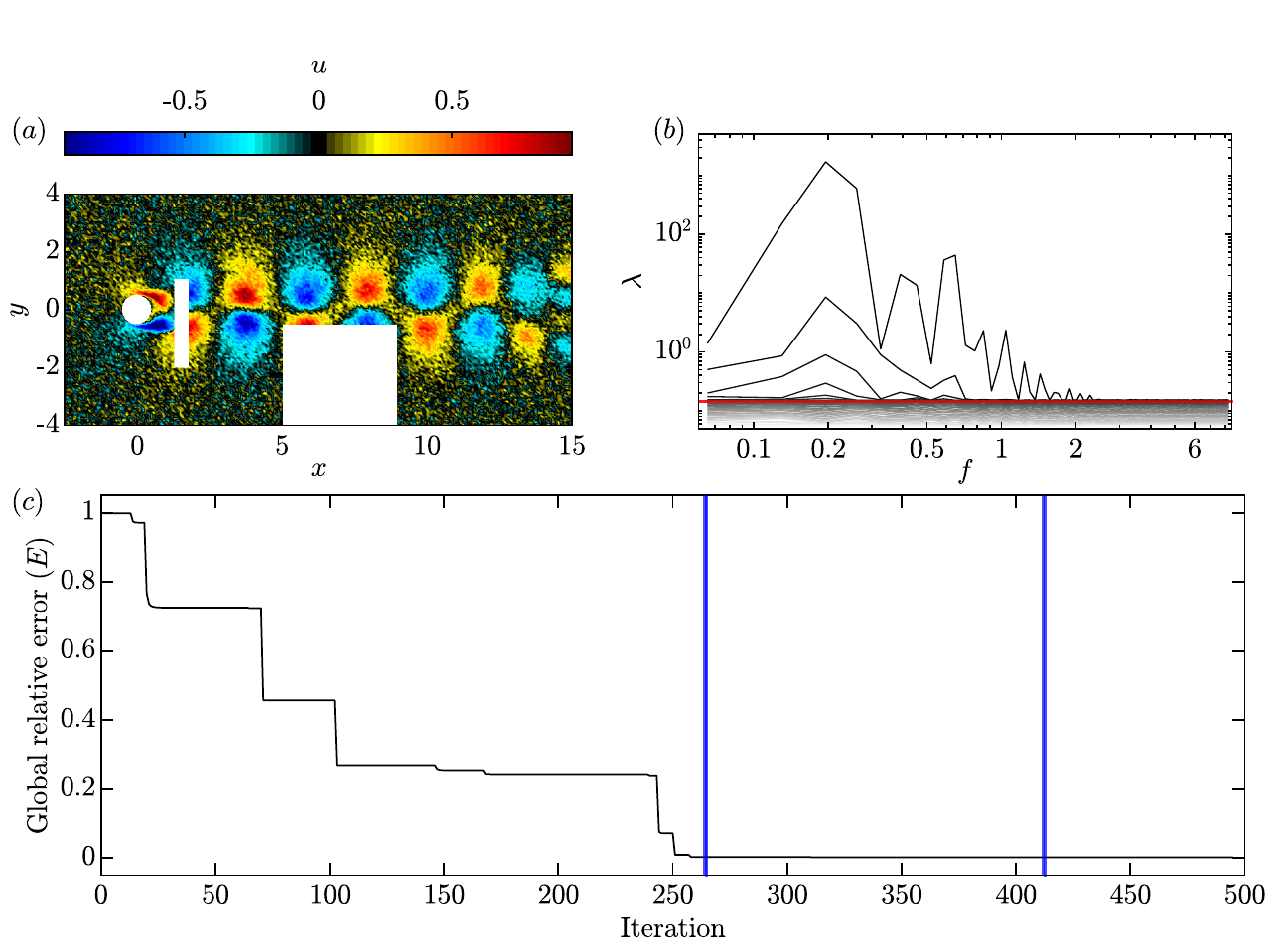}}
\caption{Gappy data reconstruction for flow with 5\% gappyness subjected to random noise with standard deviation $\sigma =0.1$: ($a$) instantaneous streamwise velocity field; ($b$) SPOD spectrum; ($c$) global relative error.  The red line in ($b$) indicates the truncation threshold, chosen slightly above the noise floor.}
\label{fig noise added}
\end{figure}

Experimental flow data such as the cavity flow data investigated in \S 4.2, inherently contains measurement noise. In \citet{nekkanti2021frequency}, we have demonstrated the capability of SPOD for denoising by rank truncation. The underlying idea is that physical structures are coherent and therefore captured by the leading SPOD modes. Many sources of measurement noise, on the other hand, are incoherent and lead to additional low eigenvalues that can conveniently be truncated. This becomes clear in figure \ref{fig noise added}($b$), where added Gaussian white noise with standard deviation, $\sigma=0.1$ leads to the formation of a distinct noise floor in the otherwise noise-free numerical cylinder data with 5\% gappyness. The same observation has been made for space-only POD by \cite{venturi2006proper} and \cite{epps2019singular}. A time instant of the streamwise velocity component, $u$, is shown in figure \ref{fig noise added} ($a$). In the gappy SPOD algorithm, we use a truncation threshold slightly above the noise floor (red line in \ref{fig noise added}($b$)) to remove noise in step (iii) before performing steps (iv)-(viii). The hard-thresholding leads to the truncation of  84\% of the modes at the lowest non-zero frequency. Frequencies above $f = 2.6$ fall below the noise floor and are removed entirely. This significant truncation in terms of number of modes, however, only amounts to 2.3\% of the total energy being removed. A positive by-product of the truncation is that the algorithm converges much faster and in fewer iterations. The total reconstruction error upon convergence is 0.19\% as compared to the $7.7 \times 10 ^{-3}\%$ of the original case in figure \ref{fig 3.2}($b$).

\begin{figure}[!tp]
\centering
{\includegraphics[trim={0cm 3.45cm 0cm 0.76cm },clip,width=1.0\textwidth]{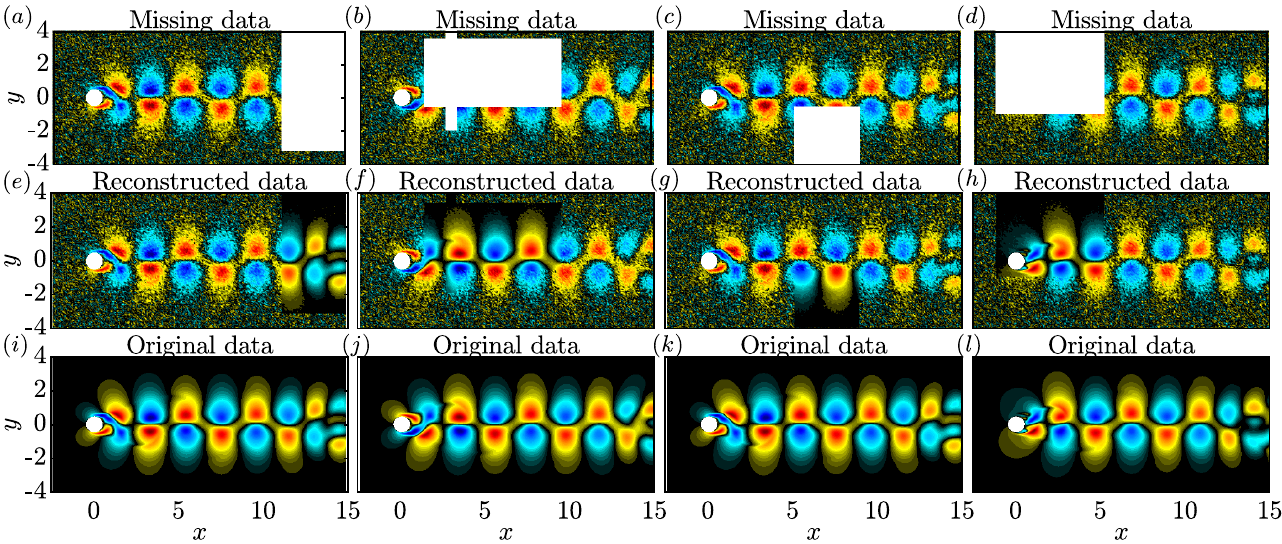}}
\caption{Same as figure \ref{fig_5_per_u_vis} with added white noise. }
\label{fig_5_per_u_vis_noise}
\end{figure}
Figure \ref{fig_5_per_u_vis_noise}  shows the instantaneous streamwise velocity component of the gappy data containing noise, reconstructed, and original flow fields, respectively. The same four time instances as in figure \ref{fig_5_per_u_vis} are selected. A visual inspection revels that the reconstructions in the missing regions are almost identical to the original data, and that the noise has been removed in large parts. The relative errors for these four snapshots are $e_i =  0.3\%$, 0.2\%, 0.15\%, and 0.17\%, respectively. Figures \ref{fig noise added} and \ref{fig_5_per_u_vis_noise} show that the algorithm not only performs well in the presence of noise, but also removes it to a large degree. 
    
\bibliography{mybibfile}

\end{document}